\documentclass[10pt,final,superscriptaddress,english,twocolumn,amssymb,aps,prb,
longbibliography]{revtex4-2}
\usepackage{graphicx}
\usepackage{natbib}
\usepackage{physics}
\usepackage{amsmath}
\usepackage{amssymb}
\usepackage{appendix}
\usepackage{bm}
\usepackage{dsfont}
\usepackage{overpic}
\usepackage[dvipsnames]{xcolor}{\huge }
\usepackage[section]{placeins}
\definecolor{darkblue}{rgb}{0,0,.65}
\definecolor{darkgreen}{rgb}{0.28,0.41,0.19}
 
\usepackage{multirow}
\usepackage{tikz}
\usepackage{nicefrac}
\usepackage[
    pdfauthor={Nils Niggemann},
  pdfstartview=FitH,
  breaklinks=true,
  bookmarks=true,
  colorlinks=true,
  anchorcolor=black,
  citecolor=blue,
  filecolor=black,
  menucolor=black,
  urlcolor=darkblue,
  linkcolor=blue,
 ]{hyperref}
\usepackage[all]{hypcap} 

\usepackage[capitalize]{cleveref}

\newcommand{\change}[1]{%
    \textcolor{red}{1}
}
\definecolor{juliared}{RGB}{202,60,50}
\definecolor{juliagreen}{RGB}{57,151,70}
\definecolor{juliapurple}{RGB}{149, 88, 178}
\definecolor{juliablue}{RGB}{64, 99, 223}
\definecolor{grey}{RGB}{180, 180, 180}

\usepackage[export]{adjustbox}
\newcommand{\subla}[1]{
{\includegraphics[width =#1\columnwidth]{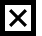}
}}

\newcommand{\sublX}{\mathchoice
  {\subla{0.03}}
  {\subla{0.03}}
  {\subla{0.018}}
  {\subla{0.015}}
}

\newcommand{\sublb}[1]{
{\includegraphics[width =#1\columnwidth]{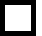}
}}

\newcommand{\sublO}{\mathchoice
  {\sublb{0.03}}
  {\sublb{0.03}}
  {\sublb{0.018}}
  {\sublb{0.015}}
}
\newcommand{\p}{\ensuremath\mathcal{P}}

\graphicspath{{images/}}
\begin{document}

\title{Classical fracton spin liquid and Hilbert space fragmentation in a 2D
spin-$1/2$ model}

\author{Nils Niggemann}
\affiliation{The Abdus Salam International Center for Theoretical Physics (ICTP), Strada
Costiera 11, I-34151 Trieste, Italy}
\affiliation{Helmholtz-Zentrum Berlin f\"ur Materialien und Energie, Hahn-Meitner Platz 1, 14109 Berlin, Germany}
\affiliation{Dahlem Center for Complex Quantum Systems and Fachbereich Physik, Freie Universit\"at Berlin, 14195 Berlin, Germany}

\author{Meghadeepa Adhikary}
\affiliation{Department of Physics and Quantum Centers in Diamond and Emerging Materials (QuCenDiEM) group, Indian Institute of Technology Madras, Chennai 600036, India}
\affiliation{SISSA, Via Bonomea 265, I-34136 Trieste, Italy}

\author{Yannik Schaden-Thillmann}
\affiliation{Helmholtz-Zentrum Berlin f\"ur Materialien und Energie, Hahn-Meitner Platz 1, 14109 Berlin, Germany}
\affiliation{Dahlem Center for Complex Quantum Systems and Fachbereich Physik, Freie Universit\"at Berlin, 14195 Berlin, Germany}

\author{Johannes Reuther}
\affiliation{Helmholtz-Zentrum Berlin f\"ur Materialien und Energie, Hahn-Meitner Platz 1, 14109 Berlin, Germany}
\affiliation{Dahlem Center for Complex Quantum Systems and Fachbereich Physik, Freie Universit\"at Berlin, 14195 Berlin, Germany}

\date{\today}

\begin{abstract}
Classical U(1) fracton spin liquids feature an extensive ground state degeneracy and follow an effective description in terms of a tensor Gauss' law where charges, so-called fractons, have restricted mobility. Here we introduce a simple spin model that realizes such a state by straightforward discretization of the higher-rank gauge theory on a square lattice. The simplicity of this construction offers direct insights into the system's fundamental fractonic properties, such as real-space fracton configurations, height-field representation of the classical ground state manifold as well as properties of local and non-local fluctuations within the fracton-free subspace. By sampling classical Ising states from the extensive ground state manifold, we show that the effective tensor Gauss' law remains intact when explicitly enforcing the spin-1/2 length constraint, demonstrating the existence of a classical Ising fracton spin liquid. However, we observe that perturbative quantum effects are insufficient to efficiently tunnel between classical ground states, leading to severe Hilbert space fragmentation which obscures fractonic quantum behavior. Specifically, by simulating the spin-1/2 quantum model with Green function Monte Carlo as a function of the Rokhsar-Kivelson potential, we find that the system supports either magnetic long-range order or a classical spin liquid. Our findings highlight the crucial role of Hilbert-space fragmentation in fractonic spin systems but also indicate ways to mitigate such effects via increasing the spin magnitude to $S=1$, investigated in a companion paper.
\end{abstract}
\maketitle

\section{Introduction}
In condensed matter physics, rich phenomena occur when the elementary degrees of freedom of a system are subject to local constraints. Well-known examples are the Hubbard model at strong coupling (no double occupancies), quantum dimer models (one dimer attached to each site) and pyrochlore spin ice (two-in-two-out spin configurations in each tetrahedron).
In the latter case, the local constraints, also called ice-rules~\cite{Moessner1998,Isakov2004}, are a direct manifestation of the system's strongly frustrated interactions.
Importantly, the dimension of the solution-space of the ice rules scales {\it exponentially} with the number of spins, a necessary condition for the emergence of a \emph{classical spin liquid}~\cite{garaninClassicalSpinLiquid1999,Moessner1998,Yan2023_1,Yan2023_2,fangClassificationClassicalSpin2024,Davier2023} at low temperatures, where the system can fluctuate freely between the available configurations. Such a state may be efficiently described by a U(1) \emph{gauge theory}, where the constraints become equivalent to a discrete form of a Gauss' law $\nabla \cdot \bm E = \rho$. Here, the charges $\rho$ (defect tetrahedra) are created in pairs by single spin flips which fractionalize and may propagate through the system. Taking quantum effects into account, the system can tunnel between different spin ice configurations, giving rise to a \emph{quantum spin liquid}, described by a compact quantum electrodynamics theory with emergent photon quasiparticles and magnetic monopoles, the latter not present in conventional continuum electrodynamics~\cite{Balents2010,Savary-2017,benton2012,Gingras2014,Pace2021,taillefumierCompetingSpinLiquids2017,Pan2016}. 

The immense potential of local constraints in spin systems to realize fascinating particles that otherwise cannot be found in nature, has motivated recent attempts to engineer even more exotic particles. One such attempt aims to realize a generalization of Gauss' law to a rank-2 gauge theory, where the electric field becomes a rank-2 tensor satisfying $\partial_\mu\partial_\nu E^{\mu\nu}=\rho$~\cite{rasmussen2016stable,Pretko2017,Pretko2017_1,Xu-2006}. Due to the conservation of charge {\it and} dipole moments in such a theory, charges, called \emph{fractons} become fully immobile, while dipoles, formed by bound pairs of charges, may move only in a direction perpendicular to their dipole moment~\cite{Nandkishore-2019,Gromov2024,Pretko-2020,You2024}. 
Fracton phases, originally discovered in lattice models extending Kitaev’s $\mathbb{Z}_2$ toric code to higher dimensions, are of significant interest for quantum information processing due to the restricted mobility of excitations, which naturally protects encoded quantum information~\cite{kitaevFaulttolerantQuantumComputation2003,Chamon-2005,Bravyi-2011,Vijay-2015,fontanaSpinorbitalKitaevModel2025}. While higher-rank U(1) theories, unlike their $\mathds{Z}_2$ counterparts, have gapless photon modes~\cite{Hermele2004,Huse2003,benton2012} and thus do not possess a robust ground state degeneracy, they can share other important properties such as violations of the eigenstate thermalization hypothesis via Hilbert space fragmentation as well as subsystem symmetries. Moreover, while $\mathds{Z}_2$ fracton liquids such as the X-Cube model, Haah's Code or Chamon's code require highly artificial multi-spin interactions that are difficult to realize~\cite{Vijay-2015,Vijay-2016,Haah-2011,Chamon-2005,Castelnovo-2012,wuU1SymmetryEnriched2023}, higher-rank spin liquids have recently been constructed at the classical level via two-body interactions enforcing generalized ice rules \cite{Benton2021}.
While it was theorized that quantum fluctuations --- in analogy to quantum spin ice --- can elevate these models to hosts quantum fracton liquids, no microscopic quantum Hamiltonian has so far been found to realize this exciting prospect~\cite{Hart2022,niggemannQuantumEffectsUnconventional2023}.

In this paper, we demonstrate that this challenge is intimately linked to the occurrence of Hilbert space fragmentation in fracton models~\cite{Sala2020,khudorozhkovHilbertSpaceFragmentation2022,Adler2024,Feng2022,stahlStrongHilbertSpace2025,Will2024}. Specifically, we introduce a particularly simple spin model with emergent fractonic properties, dubbed the \emph{spiderweb model}, which is obtained by directly discretizing the higher-rank gauge theory on a square lattice. We demonstrate that this model is ideally suited to showcase fractonic behavior on a lattice and to establish the connection between the spin model and the continuum field theory. Starting with an investigation of the classical spiderweb model in simple Gaussian approximation where the local spin-length constraint is relaxed, we develop a formalism to enumerate fracton-free states via the introduction of a height field. This enables us to identify different types of local and non-local fluctuations within the fracton-free subspace whose properties are fundamentally connected to the occurrence of Hilbert space fragmentation.

A specific focus is on the spin-$1/2$ case of the spiderweb model.
Sampling states from the exponentially large classical Ising ground state manifold, we demonstrate that the spiderweb model realizes a classical fracton liquid, evident from the presence of non-analyticities in the spin structure factor, commonly referred to as fourfold pinch points \cite{Prem2018}.
Quantum fluctuations are then introduced perturbatively, by allowing for the lowest order tunneling process between the classically degenerate states. Because of the strong Hilbert space fragmentation in the spin-$1/2$ model, the quantum dynamics induced by these tunneling processes is found to be insufficient to induce any liquid-like quantum behavior. As we demonstrate via the error-controlled Green function Monte Carlo approach, this holds in the ground-state sector, which exhibits so-called staircase order, as well as in excited sectors and even stays true at the Rokhsar-Kivelson point typically known to host a quantum spin liquid. As a result, we find the spin-$1/2$ spiderweb model to either display long-range order or to remain in a classical spin liquid state, i.e. a disordered state with an extensive number of configurations but negligible quantum fluctuations. While these results reveal the efficiency of Hilbert space fragmentation in fracton models in preventing non-trivial quantum dynamics they also suggest ways to overcome these limitations. Specifically, as we demonstrate in a companion paper~\cite{Niggemann2025b}, in the spin-1 spiderweb model, Hilbert-space fragmentation is much milder and energetically low-lying Hilbert space sectors become sufficiently large to stabilize a true quantum fracton liquid.

The remainder of the paper is structured as follows. In Sec.~\ref{sec:model} we introduce our model and discuss its general properties such as the band structure in Gaussian approximation, fracton properties in real space, conserved quantities, height-field representation of fracton-free states and fluctuator properties. In Sec.~\ref{sec:spin_half} we investigate the spin-1/2 spiderweb model where we first focus on the classical system (Sec.~\ref{sec:spin_half_classical}) and then introduce quantum dynamics (Sec.~\ref{sec:spin_half_quantum}). Our results are  summarized and discussed in \cref{sec:discussion}.

\section{Model and general properties}
\label{sec:model}
\begin{figure*}
    \centering
    \includegraphics[width = \linewidth]{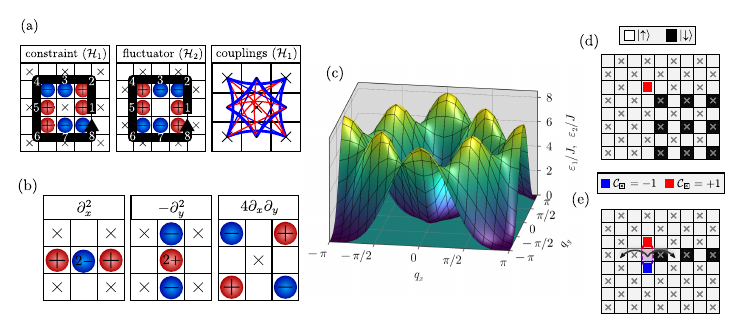}
    \caption{{\bf Definition of the spiderweb model and its elementary excitations.} (a) The ground state constraints and the fluctuators are defined on the eight-site clusters around sublattices 1 and 2, respectively, with the indicated sign structures. The couplings of the model are obtained by squaring the constraint. Red (blue) couplings are $J_{ij} = 1$ ($J_{ij} = -1$), while thick lines correspond to interactions $|J_{ij}| = 2$. (b) Derivative operators of the constraint, discretized on the square lattice. (c) Band structure of $\mathcal{H}_1$ in Gaussian approximation with a flat lower band, dispersive upper band and quartic band touching points at $\bm q =(0,0)$ and $\bm q =(\pi,\pi)$.
    (d) Isolated, immobile fracton occurring at a corner of a domain wall between two different ground state patterns. (e) A lineon, a pair of fractons, which can only move in the direction perpendicular to its dipole moment.}
    \label{fig:Overview}
\end{figure*}

We consider a model of spin-$1/2$ moments on a square lattice with sublattices 1 and 2, where we mark sublattice 1 by $\times$ in Fig.~\ref{fig:Overview}. Note that we use the convention that spins reside at the {\it centers} of the squares in Fig.~\ref{fig:Overview}. In this work, we consider particularly the case of a spin-$1/2$ system, while the $S=1$ case is the focus of a separate work~\cite{Niggemann2025b}. The Hamiltonian has three terms $\mathcal{H}=\mathcal{H}_1+\mathcal{H}_2+\mathcal{H}_3$ with
\begin{align}
\mathcal{H}_1&=\frac{J}{2}\sum_{\sublX}\mathcal{C}_{\sublX}^2,\notag\\ \mathcal{H}_2&=-J'\sum_{\sublO}\left(\mathcal{F}_{\sublO}+\mathcal{F}^\dagger_{\sublO}\right),\notag\\
\mathcal{H}_3&=\mu\sum_{\sublO}\left(\mathcal{F}^\dagger_{\sublO}\mathcal{F}_{\sublO}+\mathcal{F}_{\sublO}\mathcal{F}^\dagger_{\sublO}\right) \label{eq:Heff}
\end{align}
and
\begin{align}\label{eq:definition_cf}
\mathcal{C}_{\sublX}&=S^z_{\sublX_1}+S^z_{\sublX_2}-S^z_{\sublX_3}-S^z_{\sublX_4}+S^z_{\sublX_5}+S^z_{\sublX_6}-S^z_{\sublX_7}-S^z_{\sublX_8},\notag\\
\mathcal{F}_{\sublO}&=S^+_{\sublO_1}S^-_{\sublO_2}S^-_{\sublO_3}S^+_{\sublO_4}S^+_{\sublO_5}S^-_{\sublO_6}S^-_{\sublO_7}S^+_{\sublO_8}.
\end{align}
As indicated in \cref{fig:Overview}(a), the label $\sublX$ stands for the cluster of eight sites adjacent (along the horizontal, vertical and diagonals) to a sublattice 1 site. The cluster sites ${\sublX_1,\cdots,\sublX_8}$ are enumerated in a counterclockwise fashion starting from the site on the right of the center. Eight-site clusters around sublattice 2 sites, denoted $\sublO$, are enumerated analogously so that, e.g. $S^+_{\sublO_5}$ is a raising operator on the site left of the center of $\sublO$, see \cref{fig:Overview}(a). For conciseness of notation, we will also use the notation ${\bm S}_i$ for spin operators, where $i = 1,\dots,N_\textrm{sites}$ is a general site index not specifying its location in an eight-site cluster. Furthermore, we write ${\bm S}_{\sublX}$ (${\bm S}_{\sublO}$) to label a spin at the center of a $\sublX$ cluster ($\sublO$ cluster).

Note the special sign structure of the $S^z_i$ operators in $\mathcal{C}_{\sublX}$ which consists of alternating pairs of equal signs ${++--++\ldots}$ when moving along the cluster sites $\sublX_1,\sublX_2,\sublX_3,\ldots$. The product of raising/lowering operators $S_i^\pm$ in $\mathcal{F}_{\sublO}$ has a similar sign structure ${+--++-\ldots}$ but it is shifted by one site relative to the one in $\mathcal{C}_{\sublX}$. While $\mathcal{H}_1$ puts constraints $\mathcal{C}_{\sublX}=0$ on the ground states of the classical system with $J'=\mu=0$ for all clusters $\sublX$, the term $\mathcal{H}_2$ performs spin flips on eight site clusters $\sublO$ that add quantum dynamics to the system. Crucially, $\mathcal{C}_{\sublX}$ and $\mathcal{F}_{\sublO}$ always commute and thus $\mathcal{F}_{\sublO}$ cannot change the value of the constraint $\mathcal{C}_{\sublX}$. Furthermore, $\mathcal{H}_3$ counts the number of flippable eight-site clusters $\sublO$ and can therefore be regarded as a chemical potential for flippable clusters. In the following we will further motivate and discuss these terms starting with the classical constraint Hamiltonian $\mathcal{H}_1$.

\subsection{Constraints and Gaussian approximation}\label{sec:constraint}
Expanding the squares in $\mathcal{H}_1$ leads to a network of spin interactions up to 5th neighbors on the square lattice, which, on each cluster $\sublX$, resembles a spiderweb, see \cref{fig:Overview}(a). We therefore call our model the {\it spiderweb model}. The constraints ${\mathcal{C}_{\sublX}=0}$ are highly non-trivial for any spin length $S$. Hence, the simplest approach to understand the implications of ${\mathcal{C}_{\sublX}=0}$ is to treat individual $S_i^z$ as unconstrained classical variables (no normalization imposed). The resulting Gaussian theory can be solved exactly in momentum space in terms of the Fourier modes $S^z_m({\bm q})=\sum_{i\in m}e^{\imath {\bm q}\cdot {\bm r}_i} S_i^z$ of the spins on sublattices $m=1,2$, where ${\bm r}_i$ is the position of site $i$. Specifically, in momentum space the constraints take the simple form of a single linear equation
\begin{equation}\label{eq:constraint_q}
    \sum_{m=1}^2L^*_m (\bm q) S^z_m(\bm q) = 0,
\end{equation}
where
\begin{equation}
L_1(\bm q)=-4\sin q_x \sin q_y\;,\;
L_2(\bm q)=2(\cos q_x -\cos q_y)\;
\end{equation}
are the components of the so-called constraint vector. Crucially, the constraint vector vanishes at ${\bm q}=0$ and an expansion around this point yields a non-vanishing contribution only in {\it quadratic} order in $|{\bm q}|$. With this property, $\mathcal{H}_1$ describes an {\it algebraic} classical spin liquid according to the classification in Ref.~\cite{Yan2023_1,Yan2023_2}. Setting
\begin{equation}\label{eq:sz_versus_e}
2S^z_1({\bm q})=E^{xy}(\bm q),\quad S^z_2({\bm q})=E^{xx}(\bm q),
\end{equation}
the constraint in Eq.~(\ref{eq:constraint_q}) in lowest non-vanishing order in $|{\bm q}|$ becomes
\begin{equation}\label{eq:gauss_law}
q_\mu q_\nu E^{\mu\nu}({\bm q})=0 \text{ or } \partial_\mu \partial_\nu E^{\mu\nu}({\bm r})=0,
\end{equation}
where $E^{xx}=-E^{yy}$, $E^{xy}=E^{yx}$ and $\mu,\nu=x,y$. These long-wavelength conditions exactly correspond to the generalized charge-free Gauss' laws in momentum and direct space of a 2D electrostatic trace-less rank-2 U(1) gauge theory with an emergent matrix-valued electric field $E^{\mu\nu}$. Violations of the constraint $\partial_\mu \partial_\nu E^{\mu\nu}=\rho\neq0$ correspond to scalar type-I fracton charges $\rho$ with no mobility in real space~\cite{Pretko2017,Pretko2017_1,Prem2018,Nandkishore-2019,Pretko-2020}. The properties of $\mathcal{H}_1$ can, alternatively, be discussed by diagonalizing the $2\times2$ Hamiltonian in Gaussian approximation in momentum space. The spectrum, depicted in Fig.~\ref{fig:Overview}(c) consists of a completely flat bottom band that corresponds to the charge-free subspace and a dispersive upper band that characterizes the charges $\rho$. In accordance with a quadratically vanishing constraint vector $(L_1(\bm q),L_2(\bm q))$ at ${\bm q}=0$, the spectrum features a {\it quartic} band touching point at this momentum. We refer the reader to Appendix~\ref{app:gaussian} for further details on the Gaussian approximation.

The emergence of this higher-rank gauge constraint in our spiderweb model follows, by construction, from the choice of signs in ${\mathcal{C}_{\sublX}}$ which realize the discrete lattice versions of the derivatives $4\partial_x\partial_y$ on sublattice 1 and $\partial^2_x-\partial_y^2$ on sublattice 2, and thus represent a direct lattice discretization of Eq.~(\ref{eq:gauss_law}). For example, on a square lattice with lattice constant $a$ (which we set to $a=1$ in the following), we have $\partial^2_x f(x) \rightarrow \frac{f(x+a) - 2f(x) +f(x-a)}{a^2}$, which is visualized in Fig.~\ref{fig:Overview}(b). 

The generalized charge-free Gauss' law in Eq.~(\ref{eq:gauss_law}) is known to be associated with four-fold pinch points in the spin structure factor $\mathcal{S}({\bm q})=\frac{1}{N_\textrm{sites}}\sum_{mn}\langle S^z_m(-{\bm q})S^z_n({\bm q})\rangle$ at $T=0$~\cite{Prem2018}. These characteristic features are also revealed in our spiderweb model in self-consistent Gaussian approximation, see Fig.~\ref{fig:SpinHalfOverview}(a). Apart from the expected four-fold pinch point at ${\bm q}=(0,0)$ the same type of singularity is also obtained at ${\bm q}=(\pi,\pi)$ which represents another quartic band touching point in Fig.~\ref{fig:Overview}(c) and where an expansion of the constraint vector in quadratic order in $\bm q$ yields the same higher-rank Gauss' law as in Eq.~(\ref{eq:gauss_law}) (but with $E^{xx}\rightarrow -E^{xx}$). The existence of two momenta associated with fractonic theories follows from a symmetry of $\mathcal{H}_1$ (and also of $\mathcal{H}_2$ and $\mathcal{H}_3$) which is invariant under a $\pi$ rotation of spins around the $x$-axis on {\it one} sublattice combined with mirror reflection $x\rightarrow -x$ in direct space. For the spin structure factor this symmetry implies $\mathcal{S}(q_x,q_y)=\mathcal{S}(-q_x+\pi,q_y+\pi)$ mapping both pinch points onto each other.

\begin{figure*}
    \centering
    \includegraphics[width = \linewidth]{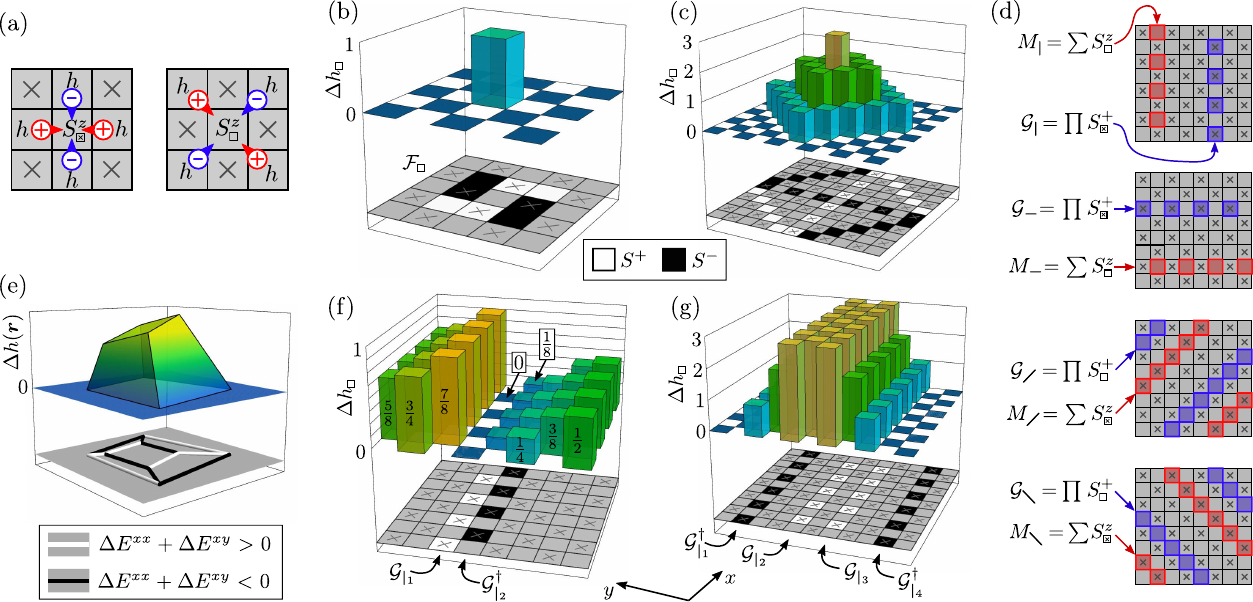}
    \caption{{\bf Height field, local and non-local fluctuators and conserved quantities.} (a) Construction of fracton-free spin states from height fields $h_{\sublO}$ as in Eq.~(\ref{eq:height_field}). The spins $S^z_{\sublX}$ (left) and $S^z_{\sublO}$ (right) are sums of the surrounding height fields $h_{\sublO}$ contributing with the depicted sign structure. (b) Unit change of height field $\Delta h_{\sublO}$ at a single site corresponding to the fluctuator $\mathcal{F}_{\sublO}$ shown at the bottom. (c) As in (b) but with a spatially extended change of height field $\Delta h_{\sublO}$. (d) Subdimensional fluctuators $\mathcal{G}_{\bm \vert}$, $\mathcal{G}_{\bm -}$, $\mathcal{G}_{\bm \diagdown}$, $\mathcal{G}_{\bm \diagup}$ (blue) and conserved magnetizations $M_{\bm \vert}$, $M_{\bm -}$, $M_{\bm \diagup}$, $M_{\bm \diagdown}$ (red). (e) Illustration of a spatially extended string-like fluctuator in the continuum limit. The addition of the height field $\Delta h({\bm r})$ creates string-like electric fields $\Delta E^{xy}$ and $\Delta E^{xx}$ along the projections of the polyhedron edges shown at the bottom. (f) Two dipole-like string fluctuators $\mathcal{G}_{{\bm \vert}_1}\mathcal{G}^\dagger_{{\bm \vert}_2}$ cannot be realized with elementary fluctuators $\mathcal{F}_{\sublO}$ as this would require fractional spin flips with $\Delta S_i^z\sim 1/L$. (g) Four string fluctuators $\mathcal{G}^\dagger_{{\bm \vert}_1}\mathcal{G}_{{\bm \vert}_2}\mathcal{G}_{{\bm \vert}_3}\mathcal{G}^\dagger_{{\bm \vert}_4}$ can be generated by integer changes of the height field $\Delta h_{\sublO}$.}
    \label{fig:fluctuator}
\end{figure*}

In Gaussian approximation where the length constraint of individual spins is relaxed, ground state spin configurations obeying ${\mathcal{C}_{\sublX}=0}$ for all eight-site clusters $\sublX$ can, in fact, be straightforwardly generated using a {\it height-field} representation. Specifically, we introduce the height field $h_{\sublO}\in\mathds{R}$ defined on the center of each $\sublO$ cluster (i.e., only on sublattice 2) and construct a spin state from $h_{\sublO}$ via
\begin{align}
S^z_{\sublX}&=h_{\sublX_1}-h_{\sublX_3}+h_{\sublX_5}-h_{\sublX_7},\notag\\
S^z_{\sublO}&=-h_{\sublO_2}+h_{\sublO_4}-h_{\sublO_6}+h_{\sublO_8}.\label{eq:height_field}
\end{align}
Here, $S^z_{\sublX}$ ($S^z_{\sublO}$) is the $z$-component of the spin in the center of a $\sublX$ ($\sublO$) cluster. Furthermore, the notations $h_{\sublX_a}$ and $h_{\sublO_a}$ are the same as in Eq.~(\ref{eq:definition_cf}) where, e.g. $h_{\sublX_1}$ is the height field on the site to the right of the center of $\sublX$ (so that $\sublX_1$ refers to a sublattice 2 site where $h_{\sublO}$ is defined). We illustrate the relation in Eq.~(\ref{eq:height_field}) between $h_{\sublO}$ and $S^z_{\sublX}$, $S^z_{\sublO}$ in Fig.~\ref{fig:fluctuator}(a). It can be easily checked that for any choice of the height field $h_{\sublO}$ the construction in Eq.~(\ref{eq:height_field}) gives rise to a spin configuration that fulfills all constraints ${\mathcal{C}_{\sublX}=0}$. This property becomes more transparent in the continuum limit $h_{\sublO}\rightarrow h({\bm r})$ where Eq.~(\ref{eq:height_field}), written in terms of the electric field [$S^z_{\sublX}\rightarrow E^{xy}({\bm r})/2$, $S^z_{\sublO}\rightarrow E^{xx}({\bm r})$, see Eq.~(\ref{eq:sz_versus_e})], becomes
\begin{align}
E^{xy}({\bm r})&=2\left(\partial_x^2-\partial_y^2\right)h({\bm r}),\notag\\
E^{xx}({\bm r})&=-4\partial_x\partial_yh({\bm r}).\label{eq:continuum_height}
\end{align}
Inserting this into the left hand side of the generalized Gauss' law in Eq.~(\ref{eq:gauss_law}), which reads as $(\partial_x^2-\partial_y^2)E^{xx}+2\partial_x\partial_y E^{xy}=0$, immediately shows the fulfillment of the ground state constraint for any height field $h({\bm r})$. On the other hand, for a system with periodic boundaries not every classical ground state can be represented by a height field, which will be discussed in more detail in the next subsection. Furthermore, if a height field exists for a given classical ground state, it is not unique because the addition of any tilted plane $h({\bm r})\rightarrow h({\bm r})+{\bm b}\cdot{\bm r}+c$ with ${\bm b}\in\mathds{R}^2$, $c\in\mathds{R}$ does not change the curvature of $h({\bm r})$ and, thus, leaves the electric field invariant. It is important to note that beyond Gaussian approximation, when treating $S_i^z$ as discrete spin-$S$ variables, a given height field $h_{\sublO}$ will usually not yield a valid spin state with $S_i^z\in\{-S,-S+1,\ldots,S\}$. Therefore, finding spin-$S$ states which fulfill all constraints $\mathcal{C}_{\sublX}=0$ is still a difficult problem, particularly for large system sizes.

In a broader context, the term $\mathcal{H}_1$ in Eq.~(\ref{eq:Heff}) can be considered as a special case in a larger class of Hamiltonians that is obtained by weighting spins $S_i^z$ on sublattice 1 with a factor $\lambda\in\mathds{R}$ in Eq.~(\ref{eq:Heff}), i.e., by substituting $S^z_{\sublX_a}\rightarrow\lambda S^z_{\sublX_a}$ in $\mathcal{H}_1$ for $a=2,4,6,8$. In Gaussian approximation, this modification preserves the bottom flat band, the quartic band touching points, the form of the Gauss' law in Eq.~(\ref{eq:gauss_law}) [but with the first equation in Eq.~(\ref{eq:sz_versus_e}) replaced by $2\lambda S^z_1({\bm q})=E^{xy}(\bm q)$] and the four-fold pinch points in the spin structure factor. On the other hand, the reweighting by $\lambda$ deforms the dispersive band $\epsilon_2({\bm q})$ in Fig.~\ref{fig:Overview}(c). Specifically, in lowest non-vanishing order in $|{\bm q}|$ the dispersive band has the form
\begin{equation}
\epsilon_2({\bm q})=\frac{J}{2}\left[(q_x^2+q_y^2)^2+4(4\lambda^2-1)q_x^2 q_y^2\right].
\end{equation}
This shows that a rotation-invariant long wavelength limit where $\epsilon_2({\bm q})\sim (q_x^2+q_y^2)^2$ requires $\lambda=1/2$, while our original model in Eq.~(\ref{eq:Heff}) with $\lambda=1$ does not have this property. Although the symmetric parameter choice $\lambda=1/2$ might appear more natural, this would lead to some serious difficulties. Specifically, the constraint ${\mathcal{C}_{\sublX}=0}$ of a model with $\lambda=1/2$ has an imbalance of weighting of spins on the two sublattices (either weighted with a prefactor $1/2$ or $1$). When considering actual discrete spin variables beyond Gaussian approximation this imbalance reduces the number of possibilities for the contributions from both sublattices to cancel each other to fulfill $\mathcal{C}_{\sublX}=0$. Furthermore, for a model with $\lambda=1/2$, a spin flip term that commutes with all $\mathcal{C}_{\sublX}$ would have a more complicated form than $\mathcal{F}_{\sublO}$ in Eq.~(\ref{eq:definition_cf}), involving more spin flip operators. To avoid such problems, throughout this paper we focus on the model as given in Eq.~(\ref{eq:Heff}) with $\lambda=1$, despite its lack of rotation invariance at long wavelengths.

We note that the recently proposed {\it honeycomb snowflake model}~\cite{Benton2021,Yan2023_1,Yan2023_2,Placke2023} is a similar 2D spin model featuring an effective rank-2 gauge constraint as in Eq.~(\ref{eq:gauss_law}). However, we consider our spiderweb model somewhat simpler since our constraint only requires eight spins, compared to twelve spins in the honeycomb snowflake model.

\subsection{Fluctuator and conserved quantities}\label{sec:fluctuator}

We continue discussing the spin flip term $\mathcal{H}_2$ that adds quantum dynamics to the system. The operators $\mathcal{F}_{\sublO}$ therein [see Eq.~(\ref{eq:definition_cf})] represent the shortest possible products of $S_i^\pm$ that are off-diagonal in a $z$-basis and that preserve the constraint, i.e., $[\mathcal{F}_{\sublO},{\mathcal{C}_{\sublX}}]=0$ for all clusters $\sublX$ and $\sublO$. An application of these {\it so-called} fluctuators $\mathcal{F}_{\sublO}$ ($\mathcal{F}^\dagger_{\sublO}$) corresponds to a local change of the height field by one unit $h_{\sublO}\rightarrow h_{\sublO}+1$ ($h_{\sublO}\rightarrow h_{\sublO}-1$) on the center of the cluster $\sublO$, see Fig.~\ref{fig:fluctuator}(b). Since any height field configuration describes a classical ground state (if the state is a valid spin-$S$ state), it is immediately clear that $\mathcal{F}_{\sublO}$ must preserve the constraints $\mathcal{C}_{\sublX}=0$.

From a physical perspective, the eight-site spin flip terms ${\mathcal{F}_{\sublO}}$ might first seem artificial, however, they naturally arise for small perturbations around $\mathcal{H}_1$. For example, $\mathcal{H}_2$ is generated in fourth order perturbation theory in transverse nearest neighbor interactions $S_i^x S_j^x+S_i^y S_j^y$, similar to the quantum dynamics generated by hexagon ring exchange terms in third order in pyrochlore quantum spin ice~\cite{Hermele2004,benton2012}. Alternatively, $\mathcal{H}_2$ is generated in eighth order perturbation theory in a transverse magnetic field $\sim S_i^x$. We also note that our eight-site fluctuator ${\mathcal{F}_{\sublO}}$ is considerably simpler than the 24 site spin flip term of the honeycomb snowflake model~\cite{Placke2023}.

The full set of fluctuators $\mathcal{F}_{\sublO}$ is complete in the sense that any {\it local} operation that carries the system from one fracton-free state to another can be expressed as a combined application of fluctuators $\mathcal{F}_{\sublO}$ on different clusters $\sublO$. A simple way to understand this is by realizing that the number of fluctuators $\mathcal{F}_{\sublO}$ (half of all lattice sites) matches the number of degrees of freedom contained in the flat band. A similar property is known from quantum spin ice where any local fluctuator can be expressed as successive applications of elementary hexagon loop moves~\cite{Hermele2004,benton2012,Bergman2008} (excluded from this property are non-local fluctuator moves that wind around periodic boundaries~\cite{Bergman2008,udagawa2024}). In quantum spin ice, these composite fluctuators have the shape of closed loops of arbitrary lengths (which are examples of Wilson loops). 
From a physical perspective, the closed loop structure of composite fluctuators means that the electric field lines their application introduces to the system cannot have end points, as these would correspond to electric charges --- a direct manifestation of the system's gauge structure which forbids the local creation of a charge. This raises a natural question of the properties of analogous loops defined via composite fluctuators formed by successive applications of elementary eight-site moves $\mathcal{F}_{\sublO}$ in our model.

In the spiderweb model, composite fluctuator moves also have the form of strings without end points, however, as we will see, these strings must have branching points, resulting in a network-like structure. To illustrate this, we consider the operation where, starting from an initial discrete and integer height field $h_{\sublO}$, we add the height field $\Delta h_{\sublO}$ shown in Fig.~\ref{fig:fluctuator}(c), top panel, i.e., we act on the configuration as $h_{\sublO}\rightarrow h_{\sublO}+\Delta h_{\sublO}$. This operation adds three layers of height fields and involves a total number of $5\times5+3\times3+1=35$ successive elementary fluctuator moves $\mathcal{F}_{\sublO}$. In terms of spin variables, the addition of this height field corresponds to the modification depicted in the bottom panel of Fig.~\ref{fig:fluctuator}(c) where spins are flipped by $S_i^+$ or $S_i^-$ along a network with a \raisebox{-0.07cm}{\rotatebox{45}{$\boxtimes$}} structure. The property of this composite fluctuator to flip spins only along a line-like structure, despite $\Delta h_{\sublO}$ being finite in an {\it area}, can be understood from the fact that $\Delta h_{\sublO}$ is a discrete version of a polyhedron with flat surfaces (in this case a pyramid). This is best illustrated in the continuum limit and ignoring the spin length constraint. For example, consider the composite fluctuator that adds the continuum height field $\Delta h({\bm r})\in \mathds{R}$ shown in Fig.~\ref{fig:fluctuator}(e), top panel, to the system. 
This height field is only non-zero in a local region where it is built from flat polygonal faces.
Since the electric fields (i.e., the $S_i^z$ components) correspond to the curvature of the height field [see Eq.~(\ref{eq:continuum_height})], the flat faces do not contribute any electric field. However, the edges of the polyhedron, where the curvature is large, give rise to a finite electric field (the orientation of the edges in the $x$-$y$ plane determine the sign and linear combination of $\Delta E^{xy}$ and $\Delta E^{xx}$ that is created). This way, the system allows for network-like fluctuator moves along the projections of the edges of any polyhedron-shaped height fields into the $x$-$y$ plane, [Fig.~\ref{fig:fluctuator}(e) bottom panel], generalizing the concept of a Wilson loop. In this picture, the spin length constraint corresponds to a condition on the curvature at the edges of the polyhedron.

\emph{Non-local fluctuators--}

While this discussion explains the structure of {\it local} fluctuator moves and how they are created by successive applications of $\mathcal{F}_{\sublO}$ there are also fluctuator moves that cannot be constructed by combining individual $\mathcal{F}_{\sublO}$. These fluctuators moves must be {\it non-local}. To understand this property it is important to note that the fluctuators $\mathcal{F}_{\sublO}$ are not completely independent since for a system with periodic boundaries the successive application of $\mathcal{F}_{\sublO}$ on all clusters $\sublO$ gives identity. Similarly, the constraints $\mathcal{C}_{\sublX}=0$ are not independent since their fulfillment on all but one cluster $\sublX$ also implies its fulfillment on this last cluster. These two relations among $\mathcal{F}_{\sublO}$ and $\mathcal{C}_{\sublX}$ mean that there must be two additional fluctuators $\mathcal{G}_1$ and $\mathcal{G}_2$ that cannot be constructed from successive applications of $\mathcal{F}_{\sublO}^{(\dagger)}$ and that preserve the constraints. It has already been noted for other flat band systems with band touching points~\cite{Bergman2008,Yan2023_2} that these `missing' fluctuators are {\it non-local}. To find $\mathcal{G}_1$ and $\mathcal{G}_2$ it is convenient to resort to a Gaussian approximation again where the fluctuator action can be mapped onto a linear algebra problem~\cite{Yan2023_1,Yan2023_2}. In this framework, a fluctuator $\mathcal{F}$ is described by a $N_{\text{sites}}$-dimensional vector where each component corresponds to a lattice site $i$. The $i$th component contains the change $\Delta S_i^z$ due to the spin flips in $\mathcal{F}$ at site $i$ and the successive application of different fluctuators corresponds to usual vector addition. Describing the constraint operators $\mathcal{C}$ analogously by $N_{\text{sites}}$-component vectors, the missing fluctuators $\mathcal{G}_1$ and $\mathcal{G}_2$ are found by the condition that the vectors for all $\mathcal{C}$, $\mathcal{F}$ as well as $\mathcal{G}_1$ and $\mathcal{G}_2$ have the rank $N_{\text{sites}}$. This yields
\begin{equation}\label{eq:g1g2}
\mathcal{G}_1=\sum_{\sublX}S_{\sublX}^+\;,\quad\mathcal{G}_2=\sum_{\sublO}S_{\sublO}^+.
\end{equation}
The operator $\mathcal{G}_1$ ($\mathcal{G}_2$) generates a homogeneous magnetization on sublattice 1 (sublattice 2) over the whole system. Since $\mathcal{G}_1$ and $\mathcal{G}_2$ cannot be constructed from single fluctuators $\mathcal{F}_{\sublO}$ the sublattice magnetizations $M_1^z=\sum_{\sublX}S_{\sublX}^z$ and $M_2^z=\sum_{\sublO}S_{\sublO}^z$ are conserved quantities. It also follows that height field representations $h_{\sublO}$ only exist for spin states with $M_1^z=M_2^z=0$. Note that $\mathcal{G}_1$ and $\mathcal{G}_2$ can create magnetization patterns with wave vectors ${\bm q}=(0,0)$ and ${\bm q}=(\pi,\pi)$ which correspond to the locations of pinch points in the spin structure factor.

Obviously, the fluctuators $\mathcal{G}_1$ and $\mathcal{G}_2$ are just two possible vector choices to complete the basis of fluctuators. Alternatively, the two missing fluctuators can be chosen from a set of non-contractible subdimensional {\it string-like} fluctuators around periodic boundaries which we denote $\mathcal{G}_{\bm \vert}$, $\mathcal{G}_{\bm -}$, $\mathcal{G}_{\bm \diagdown}$, and $\mathcal{G}_{\bm \diagup}$. These operators perform spin flips along any vertical string on sublattice 1 ($\mathcal{G}_{\bm \vert}=\prod_{\sublX\in {\bm \vert}} S_{\sublX}^+$), horizonal string on sublattice 1 ($\mathcal{G}_{\bm -}=\prod_{\sublX\in {\bm -}} S_{\sublX}^+$) or diagonal string on sublattice 2 ($\mathcal{G}_{\bm \diagdown}=\prod_{\sublO\in {\bm \diagdown}} S_{\sublO}^+$ and $\mathcal{G}_{\bm \diagup}=\prod_{\sublO\in {\bm \diagup}} S_{\sublO}^+$), see Fig.~\ref{fig:fluctuator}(d). Crucially, all these operators commute with $\mathcal{C}_{\sublX}$ and they generate finite magnetizations $M_1^z$ or $M_2^z$. Therefore, they are fluctuators that cannot be built from elementary $\mathcal{F}_{\sublO}$ moves. Similar non-compact string-like fluctuators are well-known from other flat band systems, e.g. on kagome and pyrochlore lattices~\cite{Bergman2008,udagawa2024}. Since a subdimensional (string-like) fluctuator exists for each of these strings across periodic boundaries, but only two are needed to complete the basis of fluctuators (one defined on sublattice 1 and the other on sublattice 2), it is clear that the set of all $\mathcal{G}_{\bm \vert}$, $\mathcal{G}_{\bm -}$, $\mathcal{G}_{\bm \diagdown}$, $\mathcal{G}_{\bm \diagup}$ operators and elementary $\mathcal{F}_{\sublO}$ moves must be linearly dependent. 

For example, for flat-band kagome or pyrochlore systems this means that the application of two parallel string operators is equivalent to combined elementary hexagon moves in the area between the two strings. In this last respect, however, our fractonic system behaves in a strikingly different way. Specifically, consider two parallel vertical strings on sublattice 1 denoted ${\bm \vert}_1$ and ${\bm \vert}_2$ as shown in Fig.~\ref{fig:fluctuator}(f). The linear dependence of $\mathcal{G}_{{\bm \vert}_1}$ and $\mathcal{G}_{{\bm \vert}_2}$ implies that there exists an operator $\tilde{\mathcal{F}}$ in the subspace of elementary fluctuators $\mathcal{F}_{\sublO}$ such that $\mathcal{G}_{{\bm \vert}_1}=\mathcal{G}_{{\bm \vert}_2}\tilde{\mathcal{F}}$. However, from a linear algebra perspective $\tilde{\mathcal{F}}$ corresponds to a vector with {\it non-integer} components, i.e., a non-integer change of height field $\Delta h_{\sublO}$, impossible to realize with $\mathcal{F}_{\sublO}$. As illustrated in Fig.~\ref{fig:fluctuator}(f), in the present case $\Delta h_{\sublO}$ is given by multiples of $1/L$ where $L$ is the linear system size. This implies that with enforced spin length constraints, two states related by the application of $\mathcal{G}_{{\bm \vert}_1}\mathcal{G}^\dagger_{{\bm \vert}_2}$ are in {\it different} Hilbert space sectors. This property, not present in more conventional flat band systems, is a source of Hilbert space fragmentation which will be further discussed in Appendix~\ref{app:fragmentation} and in our companion paper~\cite{Niggemann2025b}. On the other hand, a fluctuator with {\it four} strings of the form $\mathcal{G}^\dagger_{{\bm \vert}_1}\mathcal{G}_{{\bm \vert}_2}\mathcal{G}_{{\bm \vert}_3}\mathcal{G}^\dagger_{{\bm \vert}_4}$ can be generated from integer numbers of local fluctuators $\mathcal{F}_{\sublO}$ as illustrated in Fig.~\ref{fig:fluctuator}(g). The impossibility of realizing single (monopole-like) and double (dipole-like) string fluctuators using $\mathcal{F}_{\sublO}$ directly reflects the monopole and dipole conservation laws associated with the rank-2 gauge structure.

Finally, we note that in analogy to the subdimensional fluctuators $\mathcal{G}_{\bm \vert}$, $\mathcal{G}_{\bm -}$, $\mathcal{G}_{\bm \diagdown}$, $\mathcal{G}_{\bm \diagup}$, the Hamiltonian $\mathcal{H}$ also has subdimensional conserved magnetizations, which we denote $M_{\bm \vert}$, $M_{\bm -}$, $M_{\bm \diagdown}$, and $M_{\bm \diagup}$. These correspond to the $S_i^z$-sums along any vertical string on sublattice 2 ($M_{\bm \vert}=\sum_{\sublO\in {\bm \vert}} S_{\sublO}^z$), horizonal string on sublattice 2 ($M_{\bm -}=\sum_{\sublO\in {\bm -}} S_{\sublO}^z$) and diagonal string on sublattice 1 ($M_{\bm \diagdown}=\sum_{\sublX\in {\bm \diagdown}} S_{\sublX}^z$ and $M_{\bm \diagup}=\sum_{\sublX\in {\bm \diagup}} S_{\sublX}^z$), see Fig.~\ref{fig:fluctuator}(d). Note that these subdimensional conserved magnetizations are defined on the opposite sublattice to the subdimensional fluctuators. These subdimensional constants of motion can be viewed in analogy to the subsystem symmetries known from {\it gapped} fracton models~\cite{Vijay-2016}.

\subsection{Potential term}\label{sec:potential}
The last term $\mathcal{H}_3$ in Eq.~(\ref{eq:Heff}) counts the number of flippable clusters $\sublO$ and allows to tune the system between different ordered and spin liquid limits. Analogous potential terms have been included, e.g. in short-range $\mathds{Z} _2$~\cite{Moessner2001,Moessner2003,Fancelli2024} or algebraic pyrochlore U(1) spin liquids~\cite{Moessner2003,Hermele2004,benton2012,Sikora2011}. In the limit $\mu\rightarrow\infty$ ($\mu\rightarrow -\infty$) this term selects states with the minimal (maximal) number of flippable clusters which usually results in long-range order. The most interesting regime is between $\mu\sim 0$ and $\mu\sim J'$ where the competition between kinetic energy $\mathcal{H}_2$ and potential energy $\mathcal{H}_3$ may induce spin liquid behavior. This will also be the regime of primary interest in our study of the quantum spiderweb model. A special situation occurs at $J'=\mu>0$, known as the Rokhsar-Kivelson (RK) point~\cite{Rokhsar1988}, where kinetic and potential energy are perfectly balanced and an exact ground state in the form of an equal-weight superposition of all states in a given Hilbert-space sector is realized.

\section{Spin-1/2 spiderweb model}\label{sec:spin_half}
\subsection{Classical Ising model}\label{sec:spin_half_classical}
While we have demonstrated that on the Gaussian level $\mathcal{H}_1$ realizes a classical spin liquid with an emergent rank-2 gauge constraint, whether these properties survive in the spin-1/2 Ising case of $\mathcal{H}_1$ with quantized $S_i^z=\pm 1/2$ is highly non-trivial. An analogous question for the classical Ising honeycomb snowflake model was recently answered in the affirmative~\cite{Placke2023}. We first investigate this classical problem for our model by setting $J'=\mu=0$, before we add quantum dynamics in the next subsection.

\begin{figure*}
    \centering
    \includegraphics[width =\linewidth]{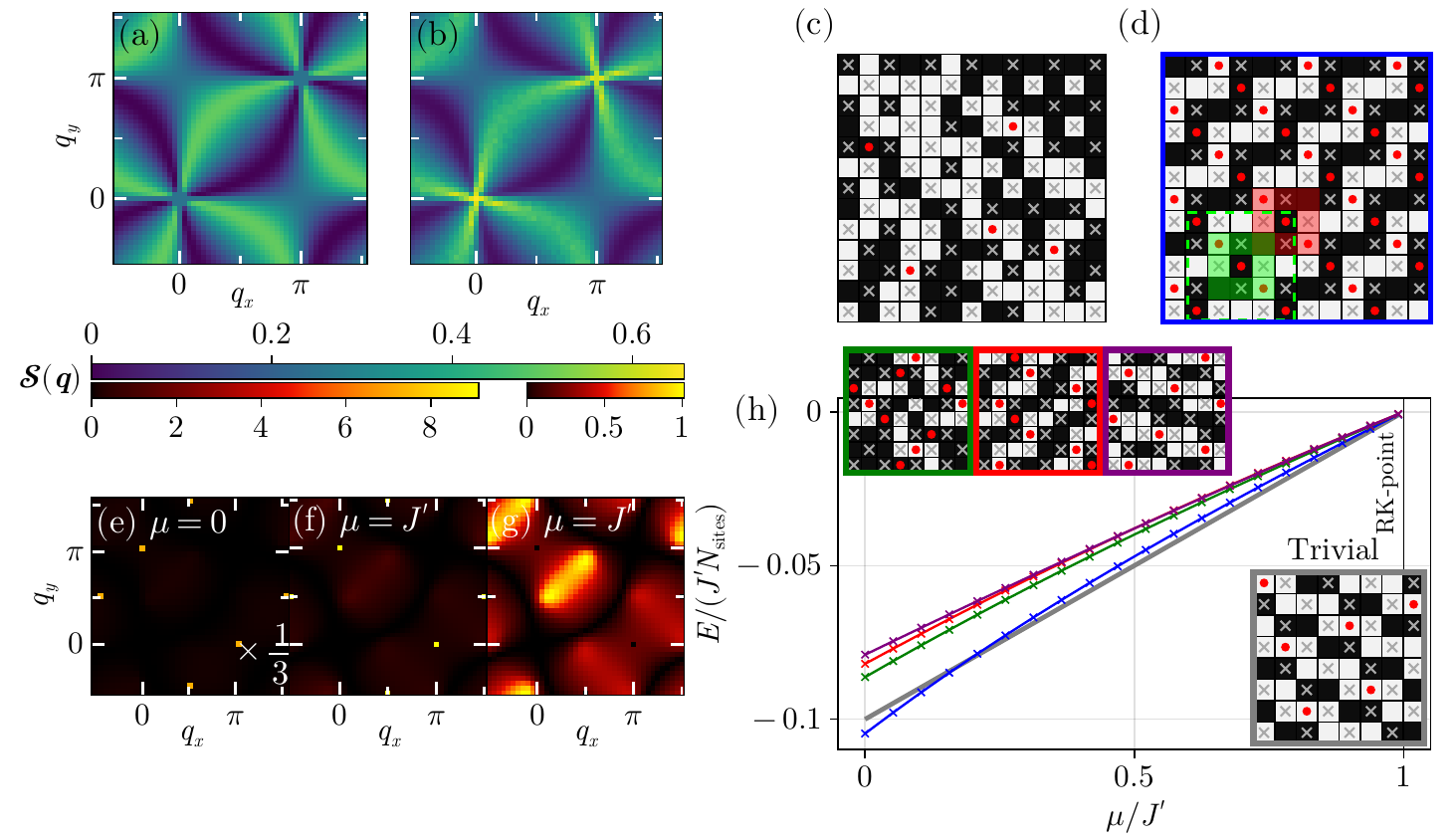}
    \caption{{\bf Properties of the spin-1/2 spiderweb model.} (a) Classical spin structure factor $\mathcal{S}({\bm q})$ in Gaussian approximation. The depicted area in momentum space corresponds to the extended Brillouin zone. (b) Spin structure factor from stochastic sampling of spin-1/2 Ising ground states of a $50\times 50$ system with open boundaries. (c) Random Ising ground state with non-overlapping flippable clusters whose centers are marked by red dots. White (black) squares denote $S_i^z=1/2$ ($S_i^z=-1/2$) sites. (d) Staircase configuration with overlapping flippable clusters that cannot be flipped simultaneously, highlighted in green and red. 
    (e) Spin structure factor from GFMC in the staircase sector of the quantum spin-1/2 model at $\mu = 0$ for a $32\times32$ system with periodic boundaries, rescaled by a factor of $1/3$. (f) Same as (e) but at the RK point $\mu=J'$ for $L=40$. (g) Same as (f) but after subtraction of the ordering wave vectors. (h) Energy as a function of $\mu$ in various sectors given by periodic initial configurations. Line colors indicate the sector shown in the insets; the blue line corresponds to the staircase state in (d). The gray line corresponds to the depicted trivial sector with non-overlapping flippable clusters. The system size is $20\times 20$ or larger, if the unit cell does not fit into the system with periodic boundaries.
    }
    \label{fig:SpinHalfOverview}
\end{figure*}

A first necessary condition for the spin-1/2 Ising spiderweb model to realize a classical spin liquid is a ground state degeneracy that scales exponentially with the number of sites. This can be proven by the identification of a spin state with a finite {\it density} of flippable clusters. Such states indeed exist; an example is the so-called {\it staircase state} shown in Fig.~\ref{fig:SpinHalfOverview}(d), a periodic configuration with a $4\times4$ unit cell where half of all clusters $\sublO$ are flippable. 
The total number of classical ground states $n=b^{N_\text{sites}}$ can be estimated via strict upper and lower bounds which gives $b=1.123 \pm 0.051$, see Appendix~\ref{app:ground_states} for details.

A way to investigate whether the exponentially many ground states of the Ising spiderweb model also give rise to an emergent charge-free rank-2 Gauss' law as in Eq.~(\ref{eq:gauss_law}), is to compute the spin structure factor $\mathcal{S}({\bm q})$ by an equal-weight sampling over large numbers of ground state configurations and check for the intactness of four-fold pinch points. An example of a generic Ising ground state is shown in Fig.~\ref{fig:SpinHalfOverview}(c). However, since the base $b$ in the exponential scaling of the number of ground states is close to one, generating such states is a difficult task, especially for large system sizes. Here, we solve this problem with advanced methods for mathematical optimization, see Appendix~\ref{app:class_sampling} for details. To make sure that these methods perform an unbiased sampling within the ground state manifold, we constrain the system even further and search for rare solutions where, additionally, a fraction of $\sim 1/6$ of all spins are fixed at random values. The spin structure factor obtained this way [see Fig.~\ref{fig:SpinHalfOverview}(b)] shows remarkable similarities with $\mathcal{S}({\bm q})$ from the Gaussian approximation in Fig.~\ref{fig:SpinHalfOverview}(a). Particularly, Fig.~\ref{fig:SpinHalfOverview}(b) reveals sharp four-fold pinch points whose width is only limited by the pixel size (i.e., by the system size, chosen as $N_{\text{sites}}\equiv L^2=50\times50$ with open boundaries). This evidences a classical fracton spin liquid in the spin-1/2 Ising spiderweb model at $T=0$.

In addition to fractonic signatures in the ground state spin correlations, the spin-1/2 Ising spiderweb model also provides a simple and intuitive understanding of fracton excitations. Specifically, an isolated fracton, i.e., a single violated constraint ${\mathcal{C}_{\sublX}\neq0}$ on one cluster $\sublX$, can be constructed as the corner of a rectangular domain wall between a ferromagnetic $S_i^z=1/2$ state and a regular pattern of three quarter $S_i^z=1/2$ and one quarter $S_i^z=-1/2$ spins (or the time-reversed $S_i^z\rightarrow -S_i^z$ version of this state), see Fig.~\ref{fig:Overview}(d). Moving such an excitation requires flipping a number of spins proportional to the linear size of the rectangular domain, which suppresses their motion for large domains. This property of fractons residing at the corners of domains, or in three dimensions on the corners of {\it membranes}, is characteristic for type-I fracton phases~\cite{Nandkishore-2019}\footnote{By contrast, in type-II fracton models, fractons are located on the corners of fractal operators, rendering all possible composite quasiparticles fully immobile.}. We note that isolated fractons can also exist on the background of irregular spin configurations (not shown). Similarly, a lineon can be formed as a dipole of fractons at the ends of a semi-infinite string of alternating $S_i^z=\pm1/2$ spins [Fig.~\ref{fig:Overview}(e)] which retains mobility in the direction perpendicular to its dipole moment, again characteristic for type-I fracton phases. We note that the string-like fluctuators discussed in Sec.~\ref{sec:fluctuator} can be viewed as the result of moving lineons along these strings.

\subsection{Quantum model}\label{sec:spin_half_quantum}
Next, we add quantum dynamics via $J'>0$ at $T=0$ where $\mathcal{H}_2$ acts on fracton-free states with $\rho=0$. A first important observation is that in most cases of randomly generated fracton-free Ising states the flippable eight site clusters $\sublO$ are sparse and non-overlapping, such that flipping one cluster neither enables nor disables other flippable clusters. This is illustrated in the representative example of Fig.~\ref{fig:SpinHalfOverview}(c) where only five non-overlapping clusters $\sublO$ are flippable. The quantum dynamics is trivial in this large majority of all Hilbert space sectors. Specifically, if in an Ising state $\ket{\psi}$ two non-overlapping clusters $\sublO$ and $\sublO'$ are flippable by $\mathcal{F}_{\sublO}$ and $\mathcal{F}_{\sublO'}$, the ground state is given by a simple product of local resonances $\sim(\mathcal{F}_{\sublO}+1)(\mathcal{F}_{\sublO'}+1)\ket{\psi}$, and each flippable cluster lowers the energy by $J' - \mu$.

The vast number of trivial Hilbert space sectors observed to be disconnected under the action of ${\mathcal{F}_{\sublO}}$ and ${\mathcal{F}^\dagger_{\sublO}}$ is peculiar and raises the question of a deeper understanding. Parts of the sectors can be explained by the conserved subdimensional magnetizations $M_{\bm \vert}$, $M_{\bm -}$, $M_{\bm \diagdown}$, and $M_{\bm \diagup}$, introduced in Sec.~\ref{sec:fluctuator} which are defined on all vertical, horizontal and diagonal strings. However, while this corresponds to a number of conserved quantities proportional to the linear system size $L$, it cannot fully explain the many observed Hilbert space sectors such as the one in Fig.~\ref{fig:SpinHalfOverview}(c).

It is tempting to try to explain the origin of the Hilbert space splitting by an analoguous effect as in more conventional two and three-dimensional spin ice-like systems~\cite{Shannon2004,Sikora2011,Pace2021}, where integer-valued winding numbers of electric flux lines define Hilbert space sectors that are topologically protected under the local Hamiltonian dynamics. At first glance, this mechanism seems possible in our system too, because one can define a new vector-valued electric field ${\bm e}$ as $e^\mu=\partial_{\nu} E^{\mu\nu}$. Rewritten in terms of this new field, the rank-2 charge-free Gauss' law $\partial_\mu\partial_\nu E^{\mu\nu}=0$ becomes a usual rank-1 Gauss' law $\partial_\mu e^\mu=0$. Since this latter condition means that field lines of $\bm e$ cannot end, one might expect that ${\bm e}$-flux around the periodic boundaries can give rise to sectors characterized by winding numbers ${\bm \phi}\in \mathds{Z}^2$. However, $\phi^x$ of a continuum system, obtained by integration over the periodic boundaries in $y$ direction, is given by
\begin{equation}
\phi^x=\oint dy\,e^x=\oint dy (\partial_x E^{xx}+\partial_y E^{xy})= \oint dy \partial_x E^{xx}\;.
\end{equation}
Using $E^{xx}=-4\partial_x\partial_y h+E_0^{xx}$ [see Eq.~(\ref{eq:continuum_height})] where $E_0^{xx}$ is a constant, ferromagnetic background on sublattice 1 that cannot be captured by the height field $h$, one obtains
\begin{equation}
\phi^x=-4\oint dy \partial_y \partial_x^2 h+\oint dy \partial_x E_0^{xx}=0\;,
\end{equation}
demonstrating that only trivial winding $\phi^x=0$ is possible. An analogous argument also proves $\phi^y=0$. We emphasize in passing that field lines in ${\bm e}$ are different from the string-like fluctuators discussed in Sec.~\ref{sec:fluctuator} and should not be confused with them. To summarize this discussion, without any obvious origin in terms of topology or symmetries, the many Hilbert space sectors of the spin-1/2 spiderweb model are an instance of Hilbert space {\it fragmentation}~\cite{Sala2020} that prevents ergodic quantum dynamics which is a common property of many fracton models~\cite{khudorozhkovHilbertSpaceFragmentation2022,Adler2024,stahlStrongHilbertSpace2025,Will2024,leeFrustrationinducedEmergentHilbert2021}. In Appendix~\ref{app:fragmentation} we provide further insights into the Hilbert space fragmentation and we prove by explicit construction of sectors that the number of fracton-free Hilbert-space sectors scales exponentially with the number of sites. 

The spin-1/2 quantum spiderweb model also has rare sectors where {\it overlapping} flippable clusters span over the whole system, allowing quantum dynamics to spread over an infinite area. In these sectors $\mathcal{H}_2$ gives rise to a true interacting many-body problem, possibly generating non-trivial quantum dynamics. We systematically searched for these sectors by constructing all periodic fracton-free spin configurations with unit cells up to 36 sites and find that the staircase state in Fig.~\ref{fig:SpinHalfOverview}(d) is the configuration with the largest possible density of flippable clusters. We note that eight degenerate staircase states exist, which are obtained by independent applications of time-reversal ($S_i^z\rightarrow -S_i^z$), translation by two lattice sites (in either $x$ or $y$ direction) and $90^\circ$ rotation in direct space. While the spin-1/2 quantum spiderweb model in the sector of the staircase state cannot be analytically solved, the quantum dynamics is unfrustrated for $J'>0$, such that sign-problem free quantum Monte-Carlo approaches become applicable. Here, we use the Green function Monte Carlo (GFMC) method explained in Appendix~\ref{app:gfmc} to investigate this system.

In Fig.~\ref{fig:SpinHalfOverview}(h) we plot the ground state energy from GFMC in the staircase sector as a function of the Rokhsar-Kivelson potential $\mu$ in the most interesting parameter regime $\mu\in[0,J']$ where kinetic and potential energy compete. We also plot the ground state energies in a few other sectors, see green, red and violet lines. These sectors result from the fluctuator dynamics acting on other periodic configurations with smaller densities of overlapping flippable clusters, see the insets in Fig.~\ref{fig:SpinHalfOverview}(h), with matching frame colors. As expected, since the staircase state maximizes the density of flippable clusters it also has the lowest energy among the sectors with overlapping flippable clusters. Furthermore, we show the energy in the sector of the state in the gray box in Fig.~\ref{fig:SpinHalfOverview}(h). This configuration with a 10-site unit cell stands out as it has the largest possible density of {\it non-overlapping} flippable clusters and thus the smallest energy of all sectors with non-overlapping flippable clusters. The quantum dynamics is trivial in that sector and corresponds to a simple product of local resonances for all flippable clusters. The crossing of the blue and gray lines at $\mu\approx0.2J'$ indicates that the ground state dynamics at $\mu\lesssim 0.2J'$ is governed by the staircase sector while it becomes trivial at $0.2J'\lesssim\mu<J'$. As expected, all curves cross at the RK point $\mu=J'$ since in this case degenerate ground states can be constructed in {\it each} sector by an equal weight superposition of all states in that sector. We refer the interested reader to Appendix~\ref{app:all_sectors} for a more involved investigation of other sectors.

To investigate the ground state quantum dynamics at small $\mu$ in more detail, we show the spin structure factor $\mathcal{S}({\bm q})$ in the staircase sector at $\mu=0$ from GFMC in Fig.~\ref{fig:SpinHalfOverview}(e) for a $32\times32$ system with periodic boundaries. The result reveals a pattern of sharp peaks at commensurate positions ${\bm q}=(-\pi/2,\pi/2)$ and ${\bm q}=(0,\pi)$ (and symmetry-related wave vectors) and only a very faint fluctuating background. These peaks are indicative of staircase magnetic order, demonstrating that despite the high density of flippable clusters, the quantum dynamics they induce is too weak to erase long-range magnetic correlations and generate a liquid-like ground state. An intuitive understanding of the weak quantum dynamics is obtained by visualizing the effects of fluctuators $\mathcal{F}_{\sublO}$ on the staircase state in direct space. As shown in Fig.~\ref{fig:SpinHalfOverview}(d) the action of $\mathcal{F}_{\sublO}$ (or $\mathcal{F}_{\sublO}^\dagger$) disables flippable clusters in a $5\times 5$ square area around it (dashed green square), but does not create any new flippable clusters. The only way to reactivate these clusters is to undo the original cluster move with $\mathcal{F}^\dagger_{\sublO}$ (or $\mathcal{F}_{\sublO}$). In this way, each fluctuator move acting on the staircase state leads to a state of more restricted dynamics, preventing the spread of collective fluctuations throughout the system.

We note in passing that the quantum dynamics from $\mathcal{H}_2$ in the staircase sector is equivalent to a typical Rydberg system of hardcore bosons~\cite{browaeysManybodyPhysicsIndividually2020,giudiciDynamicalPreparationQuantum2022,samajdarQuantumPhasesRydberg2021,zengQuantumDimerModels2025}. Viewing the staircase state as the vacuum of bosons, the application of $\mathcal{F}_{\sublO}$ or $\mathcal{F}^\dagger_{\sublO}$ on one of the flippable clusters then corresponds to the creation of a boson $b_{\sublO}^\dagger$. It follows $\mathcal{H}_2=-J'\sum_{\sublO\in\mathcal{L}} (b_{\sublO}+b_{\sublO}^\dagger)$ where $\mathcal{L}$ denotes the lattice of flippable clusters in the staircase state whose centers are marked by red dots in Fig.~\ref{fig:SpinHalfOverview}(d). Furthermore, the constraint that no overlapping clusters can be flipped is equivalent to large repulsive terms $\sim\sum_{\sublO,\sublO'\in 5\times 5} b_{\sublO}^\dagger b_{\sublO} b^\dagger_{\sublO'} b_{\sublO'}$ within the dashed green $5\times 5$ square areas in Fig.~\ref{fig:SpinHalfOverview}(d) which corresponds to the Rydberg blockade.

In Fig.~\ref{fig:SpinHalfOverview}(f) we show the spin structure factor $\mathcal{S}({\bm q})$ in the staircase sector at the RK point $\mu=J'$ for a system of size $L=40$ where the system is in an equal weight superposition of all states in the staircase sector. Interestingly, even in that case $\mathcal{S}({\bm q})$ shows sharp magnetic ordering peaks demonstrating that the long-range staircase order is still intact. However, due to the balanced effects of kinetic and potential energy, the fluctuating background signal is somewhat larger than for $\mu=0$, see also panel (g) for a version where the dominant wave vectors are subtracted. Note that all other sectors have even more restricted quantum dynamics since their flippable clusters are less dense. Thus, the overall physical situation at the RK point is as follows: The (coherent) quantum dynamics in each individual Hilbert space sector is too weak to generate a quantum spin liquid. Nevertheless, the {\it incoherent} (classical) equal probability superposition of all degenerate ground states from the different sectors restores a fracton spin liquid. This is because the spin correlations $\langle S_i^z S_j^z\rangle$ of such a state are identical to the ones of the bare spin-1/2 Ising spiderweb model (where, likewise, each configuration contributes with the same weight). These classical correlations were already shown to feature intact four-fold pinch points without any ordering peaks, see Fig.~\ref{fig:SpinHalfOverview}(a). This means that fractonic signatures in the quantum spin-1/2 spiderweb model at the RK point arise from the incoherent addition of ground states in the different sectors rather than from the quantum superposition within each sector. In other words, the spin liquid at the RK point should be considered as a classical spin liquid rather than a quantum spin liquid. This shows that strong effects of Hilbert space fragmentation may be very efficient in preventing quantum spin liquid behavior even in a flat band system tuned to the RK point -- a situation that appears ideal for quantum spin liquid.

\section{Discussion}
\label{sec:discussion}
We have introduced the spiderweb model whose simple construction enables fundamental insights into characteristic fracton properties in a lattice version of a higher-rank U(1) gauge theory. Particularly, we demonstrated the general construction of fracton-free states using a height-field representation and discussed the constraints under which the system can fluctuate between these states. Unlike in conventional U(1) gauge theories, we found that local fluctuations do not have the shape of simple closed loops, but rather form networks of strings with branching points, generalizing the concept of Wilson loops. We further confirmed that in the spin-1/2 Ising case the spiderweb model exhibits a classical spin liquid governed by the electrostatic sector of a rank-2 U(1) gauge theory. By introducing quantum dynamics via perturbative tunneling between fracton-free configurations and a Rokhsar-Kivelson potential, which increases the effect of quantum fluctuations, we uncover a rich structure of subsystem symmetries and observe that the Hilbert space fragments into an extensive number of dynamically disconnected sectors -- both hallmark features of fracton models and at stark contrast to otherwise similar spin ice models.

Known side effects of such fragmentation are glassy quantum dynamics~\cite{Chamon-2005}, which are hence expected to occur in the spiderweb model. Moreover, we find that Hilbert space fragmentation dominates the system's behavior so strongly that it obscures fractonic quantum properties and also precludes the formation of a fracton quantum spin liquid. A particularly intriguing and seemingly paradoxical consequence is that even the quantum system may realize a classical spin liquid, which we showed to be the case at the RK point. The Hilbert space is composed of extensively many quantum-disconnected sectors which are all energetically degenerate at the RK point but cannot tunnel into one another, thereby preventing the development of quantum coherence. We also investigated the strongly constrained quantum dynamics in individual sectors focusing on the largest, which we called the staircase sector. This sector displays magnetic long-range correlations even at the fine-tuned Rokhsar-Kivelson point where the ground state becomes an exact equal-weight superposition over all classical configurations. Interestingly, the quantum dynamics in this system can be mapped onto a Bose-Hubbard-type model where bosons interact repulsively via a hard-shell potential that excludes the existence of more than one boson within a certain area. In fact, this is precisely the situation that can be experimentally realized in an array of Rydberg atoms which are subject to the same type of exclusion constraint, called the Rydberg blockade~\cite{browaeysManybodyPhysicsIndividually2020,giudiciDynamicalPreparationQuantum2022,samajdarQuantumPhasesRydberg2021,zengQuantumDimerModels2025}. The effective model realized in the spin-1/2 spiderweb model at low energies is thus accessible with existing experimental techniques.

Because the spiderweb model is constructed directly from a discretization of a rank-2 U(1) gauge theory, we expect the fragmentation to be generic to other two-dimensional U(1) fracton models at spin-1/2. In those cases, the necessary tunneling processes are expected to be even more complex~\cite{Placke2023}, which would further suppress quantum dynamics and amplify the fragmentation effect.

However, for $S>1/2$ the sectors in the spiderweb model are enlarged, decreasing the impact of Hilbert space fragmentation. In our companion work~\cite{Niggemann2025b}, we show that an increase of the spin magnitude to $S=1$ is sufficient to realize a fracton quantum spin liquid with all its hallmark signatures of a higher-rank gauge theory, including emergent magnetic fields and photon excitations. We note that as $S$ tends to infinity, plaquettes (almost) always remain flippable and thus become noninteracting, eventually making the system classical.

An open question concerns how the nontrivial Hilbert-space fragmentation and fracton properties of the spin-1/2 spiderweb model are reflected in its thermodynamic behavior. This issue has been partially explored in the classical spin-1/2 Ising version of the honeycomb snowflake model, where a first-order transition into a low-temperature fracton phase was recently identified~\cite{Placke2023}. We therefore expect a similar transition to occur in our model. However, treating the full quantum system at finite temperatures poses a significant challenge, which we leave for future work.

\begin{acknowledgments}
We thank Björn Sbierski for his helpful feedback on this manuscript. We further acknowledge fruitful discussions with Subir Sachdev, Arnaud Ralko, Karlo Penc, Arnab Sen, Robin Schäfer, Daniel Lozano-Gómez, Han Yan, Yuan Wan, Alaric Sanders, Rhine
Samajdar and Marcello Dalmonte. N.~N.~and J.~R.~acknowledge support from the Deutsche
Forschungsgemeinschaft (DFG, German Research Foundation), within Project-ID 277101999 CRC 183 (Project
A04). N.~N.~ further acknowledges funding from the European Research
Council (ERC) under the European Union’s Horizon ERC-2022-COG Grant with Project No. 101087692.
M.~A.~ acknowledges funding from the PNRR MUR project PE0000023-NQSTI.
We acknowledge the use of the JUWELS cluster at the Forschungszentrum J\"ulich and the Noctua2 cluster at the Paderborn Center for Parallel Computing (PC$^2$).
\end{acknowledgments}
\section*{Data availability}
The data that support the findings of this article are openly available \cite{zenodo_repository}.
\bibliography{bib}

\begin{appendices}
\section{Gaussian approximation of the classical model}\label{app:gaussian}
We start by writing the ground state constraint $\mathcal{C}_{\sublX}=\sum_{a=1}^8 \sigma_{\sublX_a} S^z_{\sublX_a} = 0$ defined via $\mathcal{H}_1$ in \cref{eq:Heff} where $\sigma_{\sublX_a} \in \{+1,-1\}$ [see \cref{fig:Overview}(a)] in momentum space as 
\begin{equation}\label{eq:constraint_L}
    \sum_{m=1}^2L^*_m (\bm q) S^z_m(\bm q) = 0,
\end{equation}
where
\begin{align}
L_1(\bm q)&=\sum_{a=2,4,6,8}\sigma_{\sublX_a}e^{\imath \bm q \cdot{\bm r}_{\sublX_a}}=-4\sin q_x \sin q_y\;,\\
L_2(\bm q)&=\sum_{a=1,3,5,7}\sigma_{\sublX_a}e^{\imath \bm q \cdot{\bm r}_{\sublX_a}}=2(\cos q_x -\cos q_y)\;.\label{eq:constraint_L2}
\end{align}
Here, ${\bm r}_{\sublX_a}$ is the real space position of a site $i$, located relative to the center of a cluster $\sublX$ in the direction specified by the number $a=1,\ldots,8$ as shown in \cref{fig:Overview}(a).

The constraint in Eq.~(\ref{eq:constraint_L}) implies that the Fourier components $(S_1^z(\bm q),S_2^z(\bm q))$ of the ground state spin configurations need to be orthogonal to the vector $(L_1(\bm q),L_2(\bm q))$ at each point in reciprocal space. 
The spin-1/2 normalization constraint $(S^z_i)^2 = 1/4$ is in practice difficult to enforce in this momentum space description and is instead approximated by an averaged one \begin{equation}
    \sum_{i} (S^z_i)^2 = \frac{2}{N_\text{sites}}\sum_{\bm q,m} S^z_m(\bm q)S^z_m(-\bm q)= \frac{N_\textrm{sites}}{4}. \label{eq:avgConstraint}
\end{equation}
With this approximation the model becomes a classical Gaussian theory subject to a single (global) constraint. The spin structure factor $ \mathcal{S}(\bm q)$ can then be obtained by projecting out all Fourier components `parallel' to $(L_1(\bm q),L_2(\bm q))$:  
\begin{align}\label{eq:ssf_class_gaussian}
    \mathcal{S}(\bm q) &\equiv  \frac{1}{N_\textrm{sites}}\sum_{mn} \langle S^z_m(-\bm q)S^z_n(\bm q)\rangle \\
    &\propto\sum_{mn}\left( \delta_{mn} - \frac{L_m(\bm q)L_n(\bm q)}{ L_1^2(\bm q)+L_2^2(\bm q)}\right)\\
    &=\frac{(c_x-c_y+2s_x s_y)^2}{(c_x-c_y)^2+4 s_x^2 s_y^2}
\end{align}
where we use the shorthand notation $c_\mu=\cos q_\mu$, $s_\mu=\sin q_\mu$ with $\mu=x,y$. The missing proportionality constant in the second line of Eq.~(\ref{eq:ssf_class_gaussian}) needs to be adjusted such that the sum rule in Eq.~(\ref{eq:avgConstraint}) is fulfilled. A known property of this description is that the spin structure factor is non-analytic at points in momentum space where $( L_1 (\bm q),L_2(\bm q)) = 0$ giving rise to pinch points indicating an emergent gauge theory~\cite{Benton2021}. In our case $( L_1 (\bm q),L_2(\bm q)) = 0$ for $\bm q =(0,0)$ and $\bm q =(\pi,\pi)$. Particularly, since also the {\it linear} terms in $\bm q$ vanish at these two points, i.e., $\partial_\mu( L_1 (\bm q),L_2(\bm q)) = 0$ for $\mu=x,y$, the pinch points are four-fold [see Fig.~\ref{fig:SpinHalfOverview}(a)] as is characteristic for a rank-2 U(1) gauge theory~\cite{Prem2018}.

An alternative way of discussing these properties is by diagonalizing $\mathcal{H}_1$ in the space of the Fourier components $S_1^z(\bm q)$ and $S_2^z(\bm q)$. Using that $\mathcal{H}_1$ can be written as
\begin{equation}\label{eq:hbands}
\mathcal{H}_1=\frac{J}{N_\textrm{sites}}\sum_{\bm q,m,n}S_m^z(-\bm q) L_m(-\bm q)L_n(\bm q)S_n^z(\bm q)
\end{equation}
this amounts to diagonalizing the $2\times 2$ matrix
\begin{equation}
H_{mn}=\frac{J}{2}L_m(-\bm q)L_n(\bm q)\;,
\end{equation}
which, due to its simple projector-like form, has eigenvalues $\epsilon_1=0$ and $\epsilon_2=J(L_1^2(\bm q)+L_2^2(\bm q)/2$. These eigenvalues form a bottom flat band and a higher dispersive band with band touching point at $\bm q =(0,0)$ and $\bm q =(\pi,\pi)$. Particularly, as a consequence of the quadratic behavior of $L_1(\bm q)$ and $L_2(\bm q)$ around these points, the band touching is of {\it quartic} type.

\section{Hilbert space fragmentation}\label{app:fragmentation}
The repeated action of fluctuators $\mathcal{F}_{\sublO}$ and $\mathcal{F}_{\sublO}^\dagger$ does not allow to visit all spin configurations that fulfill the constraints $\mathcal{C}_{\sublX}=0$, but splits up the fracton-free Hilbert space into many sectors disconnected under $\mathcal{F}_{\sublO}$ and $\mathcal{F}_{\sublO}^\dagger$. Here, we provide insights into the structure of these sectors and prove that the spin-1/2 spiderweb model exhibits exponentially many of such sectors, giving rise to a severe Hilbert space fragmentation.

\begin{figure}
    \centering
    \includegraphics[width = 0.9\linewidth]{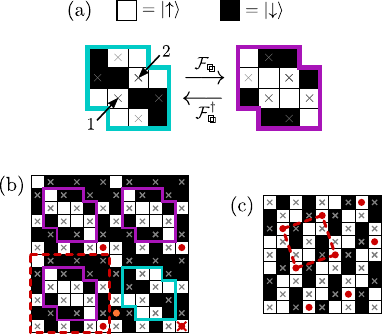}
    \caption{(a) Action of the simultaneous double fluctuator $\mathcal{F}_{\includegraphics[scale=0.2]{double_move.pdf}}$. Note that $\mathcal{F}_{\includegraphics[scale=0.2]{double_move.pdf}}$ leaves the two sites 1 and 2 unchanged. (b) Periodic $6\times6$ spin configuration in which any application of $\mathcal{F}_{\includegraphics[scale=0.2]{double_move.pdf}}^\dagger$ creates a new Hilbert space sector. (c) Spin configuration to estimate a lower bound for the exponential scaling of the number of classical ground states.}\label{fig:construction_sectors}
\end{figure}

Our proof considers the composite fluctuator $\mathcal{F}_{\includegraphics[scale=0.2]{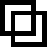}}$ illustrated in Fig.~\ref{fig:construction_sectors}(a) that acts on a 14-site cluster. This operator flips 12 sites of the cluster while leaving two center sites [labeled 1 and 2 in Fig.~\ref{fig:construction_sectors}(a)] unchanged. Crucially, we define $\mathcal{F}_{\includegraphics[scale=0.2]{double_move.pdf}}$ such that it only acts on states where these center spins are equal (while $\mathcal{F}_{\includegraphics[scale=0.2]{double_move.pdf}}$ annihilates the state otherwise). In the depicted case they are in a $|{\uparrow}\rangle$ configuration, however, the same argument also holds for $|{\downarrow}\rangle$ states. The action of $\mathcal{F}_{\includegraphics[scale=0.2]{double_move.pdf}}$ corresponds to the combination of two elementary eight-site fluctuators $\mathcal{F}_{\sublO}$ and $\mathcal{F}_{\sublO'}$ on overlapping clusters that are shifted in a diagonal direction with respect to each other. Therefore, $\mathcal{F}_{\includegraphics[scale=0.2]{double_move.pdf}}$ commutes with all $\mathcal{C}_{\sublX}$ which implies that $\mathcal{F}_{\includegraphics[scale=0.2]{double_move.pdf}}$ switches between different fracton-free configurations. Most importantly, the restriction that $\mathcal{F}_{\includegraphics[scale=0.2]{double_move.pdf}}$ only acts on states where the two center spins are equal means that {\it no sequential action} of elementary fluctuators $\mathcal{F}_{\sublO}\mathcal{F}_{\sublO'}$ or $\mathcal{F}_{\sublO'}\mathcal{F}_{\sublO}$ can reproduce $\mathcal{F}_{\includegraphics[scale=0.2]{double_move.pdf}}$ since this always annihilates the state. This means that two fracton-free configurations connected by $\mathcal{F}_{\includegraphics[scale=0.2]{double_move.pdf}}$ lie in different Hilbert space sectors of the quantum spin-1/2 spiderweb model, at least when considering systems consisting only of the 14 sites on which $\mathcal{F}_{\includegraphics[scale=0.2]{double_move.pdf}}$ acts.

Next, we investigate situations where the 14-site cluster of Fig.~\ref{fig:construction_sectors}(a) is embedded in a larger system. In this case it is {\it a priori} unclear whether $\mathcal{F}_{\includegraphics[scale=0.2]{double_move.pdf}}$ still creates a new Hilbert space sector. This is because the sequential applications of $\mathcal{F}_{\sublO}$ and $\mathcal{F}_{\sublO'}$ may become possible when performed in a suitably chosen sequence together with applications of fluctuators on other, surrounding eight-site clusters. In this case $\mathcal{F}_{\includegraphics[scale=0.2]{double_move.pdf}}$ would not create a new sector. However, this possibility can be excluded in specific states. For example, let us consider the configuration in Fig.~\ref{fig:construction_sectors}(b) which is a periodic pattern with a $6\times6$ unit cell, indicated by the dashed red square. Each unit cell has one eight-site cluster flippable by $\mathcal{F}_{\sublO}^\dagger$ (their centers are marked by red dots) and one 14-site cluster flippable by $\mathcal{F}_{\includegraphics[scale=0.2]{double_move.pdf}}^\dagger$ (marked by violet lines). More specifically, the application of $\mathcal{F}_{\includegraphics[scale=0.2]{double_move.pdf}}^\dagger$ is illustrated in the lower right $6\times6$ unit cell in Fig.~\ref{fig:construction_sectors}(b) which creates one additional flippable eight-site cluster (orange dot) but inactivates an existing one (crossed red dot). The $\mathcal{F}_{\sublO}$-dynamics in both states (i.e., before and after the application of $\mathcal{F}_{\includegraphics[scale=0.2]{double_move.pdf}}^\dagger$) is very restricted and only occurs on single disconnected flippable eight-site clusters or on pairs of overlapping eight-site clusters [nearby red and orange dots in Fig.~\ref{fig:construction_sectors}(b)] such that non-trivial collective phenomena cannot occur in these states. Particularly, sequential applications of $\mathcal{F}_{\sublO}$ and $\mathcal{F}_{\sublO'}$ in a 14-site cluster in combination with surrounding fluctuator moves are not possible. It follows that the application of $\mathcal{F}_{\includegraphics[scale=0.2]{double_move.pdf}}^\dagger$ in any unit cell creates a new sector, such that a lower bound for the number of sectors can be estimated as $2^{N/36}$, proving the exponential scaling of the number of sectors with system size.

\section{Number of classical ground states}\label{app:ground_states}
Here, we estimate the number of classical states $n$ that fulfill all the ground state constraints $\mathcal{C}_{\sublX}=0$ in the spin-1/2 spiderweb model. Specifically, considering a system with $N_{\text{sites}}$ spins we are interested in the value of the base $b$ of the exponential scaling $n=b^{N_{\text{sites}}}$ of such states in the thermodynamic limit. To estimate an upper bound of $b$, we consider quadratic systems with $N_{\text{sites}}=L\times L$ sites and periodic boundary conditions and determine $n$ for various sizes $L$ via brute force generation of all possible states. With reasonable numerical effort this can be done up to $L=12$. The obtained values are listed in Table~\ref{tab:num_GS} (note that $L$ must be even). For each fixed system size $L$, the base follows from $b=n^{1/L^2}$. As can be seen in Table~\ref{tab:num_GS}, the base $b$ decreases with $L$. Hence, the obtained $b$ for the largest considered system size can be taken as an upper bound for approximating the thermodynamic limit.

\begin{table}
\begin{tabular}{|c|c|c|}
\hline
& Number of states $n$ & $b=n^{1/L^2}$\\\hline
$L=4$ & 216 & 1.399\\
$L=6$ & 5912 & 1.273\\
$L=8$ & 350872 & 1.221\\
$L=10$ & 37403668 & 1.190\\
$L=12$ & 10103561614 & 1.173\\
\hline
\end{tabular}
\caption{Total number of states $n$ that fulfill the classical ground state constraints $\mathcal{C}_{\sublX}=0$ for a system with $N_{\text{sites}}=L\times L$ sites and periodic boundary conditions for different $L$. The value $b=n^{1/L^2}$ in the third column is the base for the respective system size $L$ from which the total number of classical ground states follows by $n=b^{N_{\text{sites}}}$.}\label{tab:num_GS}
\end{table}

A lower bound of $b$ can be determined from the state in Fig.~\ref{fig:construction_sectors}(c). This configuration has a square-shaped $\sqrt{10}\times\sqrt{10}$ unit cell (red dashed square) and one flippable cluster per unit cell, the centers of which are marked by red dots. Since flippable clusters correspond to the eight sites around the red dots, all flippable clusters of the state in Fig.~\ref{fig:construction_sectors}(c) are {\it non-overlapping}. Importantly, this configuration realizes the largest possible density of non-overlapping flippable clusters.
Since each flippable cluster can be in two configurations, these flips can generate $2^{N_{\text{sites}}/10}$ different classical ground states. This yields a lower bound for $b$ given by $b=2^{1/10}=1.072$. Taking the mean values between these upper and lower bounds we obtain
\begin{equation}
b = 1.123 \pm 0.051\label{eq:hilbert_space_size}
\end{equation}
as a rough estimate for $b$, where the error margins range up (down) to the upper (lower) bounds.

It is worth noting that estimates for $b$ from usual Pauling counting arguments are surprisingly inaccurate. Such an argument would approximate the total number of classical ground states $n$ as $n\approx2^{N_{\text{sites}}} k^{N_{\text{sites}}/2}$ where $2^{N_{\text{sites}}}$ is the total number of possible Ising configurations in a system with $N_{\text{sites}}$ spins, $N_{\text{sites}}/2$ is the number of constraints and $k$ is the ratio between the number of all configurations with $\mathcal{C}_{\sublX}=0$ for one cluster $\sublX$ and the total number of $2^8$ Ising configurations in one cluster. 
Since one cluster $\sublX$ has $\binom{8}{4}=70$ states that fulfill $\mathcal{C}_{\sublX}=0$, this argument approximates $b$ as $b\approx\sqrt{35/32}= 1.046$, which, however, is even smaller than our lower bound in Eq.~(\ref{eq:hilbert_space_size}).

\section{Sampling of classical ground states}
\label{app:class_sampling}
The treatment of classical spin liquids via the Gaussian approximation is usually observed to describe the qualitative nature of a spin model well. However, it is not at all obvious whether its predictions hold in the case of Ising spins.
In order to establish fractonic behavior in this case, we need to explicitly respect the local constraint $(S^z_i) ^2 = 1/4$, and perform a uniform sampling among all configurations which fulfill the ground state constraint $\mathcal{H}_1 = 0$ in \cref{eq:Heff}, a highly nontrivial challenge. 
A natural approach to this problem is to employ a classical Monte Carlo simulation, cooling a system down to temperatures near zero, until a true ground state is found (whose validity may be verified quickly). While theoretically possible, this approach is severely limited due to the extremely limited mobility of fractons: Since a single fracton cannot be moved without creating even more fractons, traditional Monte Carlo update schemes involving single spin flips are inapplicable, as the acceptance rate of any move vanishes at low temperatures, leading to glassy dynamics. This problem was explored in more detail in Ref.~\cite{Placke2023}, where advanced update schemes have been proposed that make use of local moves similar to the one shown in \cref{fig:Overview}(a), as well as lineon-like propagations similar to \cref{fig:Overview}(e). While such an approach was shown to reach zero-charge ground states, it remains numerically challenging to sample ground states as legal lineon moves become less probable with increasing system size.
Instead, here we attempt to construct valid ground states directly.

To allow a sampling with a controllable bias, we may rephrase our problem in the domain of mathematical optimization, a field which aims to find solutions of constrained problems.
The ground state constraints can be posed as a linear system of equations for the spins $S^z_1,\dots S^z_{N_\textrm{sites}}$
\begin{equation}
    \sum_j \mathcal{C}_{ij} S^z_j = 0
\end{equation}
where $\mathcal{C}_{ij}$ is a constraint matrix of dimension $N_\textrm{sites}\times N_\textrm{sites}/2$. For continuous spins without local length constraints, this can be algebraically solved, however the spin length constraint $(S^z_i)^2 = 1/4$ requires more sophisticated tooling such as \emph{mixed integer programming} solvers. To employ these techniques we only need to provide the set of constraints as well as an empty objective function (as here we do not intend to optimize anything).
It is evident that this approach will still be highly biased, as the constraint solver certainly does not choose uniformly among the feasible solutions. 

We overcome this bias systematically by adding additional random constraints with the goal of restricting the solution space to a few, or even a single solution. These constraints can be arbitrary. Here, we randomly select a fraction of all spins and initialize them by a random value. The size of this fraction controls the bias: If too many constraints are imposed, there will usually be no solution, leading to inefficient sampling. On the other hand, if too few spins are chosen then there are many solutions and the sampling is not guaranteed to be unbiased. Ideally, we want to choose this fraction such that the system only has a single solution. Here, we chose a large fraction of $N_\textrm{sites}/6$ randomly initialized spins, which is quite overconstrained, and typically leads to no solutions. To address the tremendously difficult task of finding a solution in this highly constrained space, we use the commercial software \emph{Gurobi} \cite{gurobi} which we find to be significantly faster (though otherwise equally applicable) than non-commercial alternatives, such as SCP \cite{BolusaniEtal2024OO}.
By repeating this process many times we eventually accumulate enough solutions that are virtually uncorrelated among each other. The result is shown in \cref{fig:SpinHalfOverview}(b) and shows a structure factor surprisingly close to the one found in the Gaussian approximation.

\section{Implementation details of Green function Monte Carlo}\label{app:gfmc}
In the following, we detail the Green function Monte Carlo (GFMC) approach. Rather than repeating the rigorous derivation provided in Refs.~\cite{Buonaura1998,becca2017quantum,Trivedi1990,Hetherington1984}, the focus of this section is to discuss the approach in more intuitive terms using the concrete context of the spiderweb model.

In the present work, we are interested in the properties of the ground state of the Hamiltonian in \cref{eq:Heff}. The exact solution to this problem requires the representation of this ground state $\ket{\psi}$ in terms of an exponential number of spin configurations, here labeled as $\ket{x} = \ket{S^z_1,\dots,S^z_{N_\textrm{sites}}}$ for conciseness of notation,
\begin{equation}
    \ket{\psi} = \sum_x \psi(x) \ket{x} .
\end{equation}
While $\psi(x)\equiv \braket{x}{\psi}$, is conventionally interpreted as the component of a $2^{N_\textrm{sites}}$ dimensional vector, here we emphasize the equivalent viewpoint of $\psi(x)$ as a function which maps a spin configuration onto a corresponding amplitude.

GFMC manages this exponential scaling of the Hilbert space by performing statistical sampling of observables via a Markov chain.
Properties of the ground state $\ket{\psi}$ of a Hamiltonian $\mathcal{H}$ can be sampled via a projective approach on any trial wavefunction $\ket{\psi_T}$ with nonzero overlap to the ground state $\ket{\psi}$
\begin{equation}
    \ket{\psi} \sim \lim_{\mathcal{P}\rightarrow \infty} (\Lambda  - \mathcal{H})^{\p} \ket{\psi_T}. \label{eq:projection}
\end{equation}
Here, $\Lambda \geq0$ is a uniform energy shift that can be used to ensure positive definiteness. This procedure converges to the ground state exponentially fast in the projection order $\mathcal{P}$ if the system is gapped (which holds for any finite system size).
If the projection matrix $\mathcal{G} \equiv \Lambda - \mathcal{H}$ has no negative elements, one may interpret its elements as transition probabilities and use stochastic approaches to evaluate \cref{eq:projection}.
In this case, the Perron–Frobenius theorem states that its dominant eigenvalue, i.e. the eigenvalue with the largest magnitude, is given by a positive real number, while the corresponding eigenvector contains only positive numbers. This is ensured by the conditions $\Lambda \geq \max_x \mathcal{H}_{xx}$, and $\mathcal{H}_{xx'}\leq0$ for $\ket{x'} \neq \ket{x}$.

To illustrate how the formalism works, we postpone the general procedure and first consider an instructive special case in which $\mathcal{G}_{xx'}$ is equal to a doubly stochastic transition matrix $p_{xx'} =p_{x'x} \geq 0$ which fulfills the property
\begin{align}
    \sum_{x'} p_{x x'} &= 1. \label{eq:MarkovMatrix}
\end{align}
This is for instance realized in a trivial sector with $N_{\sublO}$ non-overlapping flippable clusters such that none of the moves affect each other. For simplicity, we take $\mu=\Lambda=0$ and $J' = -1$, after which we can easily identify $\mathcal{G}_{x x'} = p_{x x'} N_{\sublO}$. 

At each step $\ket{x}$ of the random walk, (starting from a selected initial configuration $\ket{x_1}$), we select one of the configurations $\ket{x'}$ with probability $p_{xx'}$. This defines a Markov chain of configurations $\ket{x_n}$, where $n$ labels the step index. After sufficiently many steps, the probability $\Phi^\textrm{Eq}(x)$ to be in a particular state $x$ is given by an equilibrium condition, which is independent of the initial state
\cite{Hetherington1984}
\begin{equation}
    \sum_{x'} p_{x x'} \Phi^\textrm{Eq}(x') = \Phi^\textrm{Eq}(x).
\end{equation}
Interpreting the above as an eigenvalue equation, we see that the equilibrium probability distribution $\Phi^\textrm{Eq}(x)$ is proportional to an eigenvector of $p_{x x'}$ with eigenvalue $1$ and thus also an eigenstate of $\mathcal{H}$. It can be easily seen that this is fulfilled by the eigenvector given as $\Phi^\textrm{Eq}(x) = 1 , \ \forall x$. Moreover, for Markov matrices fulfilling \cref{eq:MarkovMatrix}, the spectrum of $p_{x x'}$ is bounded such that all other eigenvalues $\lambda_i \leq 1$ \cite{Hetherington1984}.
Since the largest eigenvalue of $p_{x x'}$ corresponds to the minimum eigenvalue of $\mathcal{H}_{x x'}$, we can identify $\psi(x) = \sqrt{\Phi^\textrm{Eq}(x)}$. 
Here, it is instructive to point out an observation: The ground state eigenvector of $\mathcal{H}$ has only non-negative elements. This property is rather special: Hamiltonians whose ground state contains arbitrary phases (in the computational basis), usually cannot be represented by a stochastic matrix. This problem is another possible definition of the so-called \emph{sign-problem}. 

Presently, the probability distribution of configurations of the Markov chain $\ket{x_n}$ is equal to the probability distribution of the ground state wavefunction. As a result we may determine expectation values of any observable $\mathcal{O}$ in the ground state by averaging its \emph{local estimator} $\mathcal{O}_L(x) = \mel{\psi}{\mathcal{O}}{x}/\braket{x}{\psi}$ over a large enough number $N$ of classical configurations $\ket{x_n}$, drawn from an equilibrated Markov chain
\begin{align}
    \langle \mathcal O \rangle &= 
    \frac{\sum_x |\psi(x)|^2 \mathcal{O}_L(x)}{\sum_x |\psi(x)|^2}\\
    &\approx 
    \frac{1}{N}\sum^N_n \mel{x_n}{\mathcal{O}}{x_n}.
\end{align}
In the second line, we used the fact that $\Phi^\textrm{Eq}(x) = 1$. 

We now discuss more general cases in which $\mathcal{G}_{x x'} \not\propto p_{x x'}$ (while still being positive!). In this case, we may still define a random walk by defining normalized probabilities as
\begin{equation}
    p_{x x'} = \frac{\mathcal{G}_{x x'}}{\sum_{x'}\mathcal{G}_{x x'} } \equiv \frac{\mathcal{G}_{x x'}}{w_x} .\label{eq:prob_weights}
\end{equation}
Clearly, we now have for the ground state $\psi(x) \not \propto \Phi^\textrm{Eq}(x)$,
however, the information about $\braket{x}{\psi}$ is contained in the joint probability distribution for $(w_{x_n}, x_n )$ \cite{Hetherington1984}.

To measure an observable, we need to employ the projection. Concretely, our trial state is given by $\psi_T(x) = 1$. Formally, the projection is achieved by replacing $\ket{\psi} \rightarrow \mathcal{G}^{\p} \ket{\psi_T}$ in $\expval{\mathcal{O}} = \mel{\psi}{\mathcal{O}}{\psi}$ to obtain
\begin{align}
    \langle \mathcal O \rangle &= \lim_{\p \rightarrow \infty} \frac{\mel{\psi_T}{\mathcal{G}^\p\mathcal O\mathcal{G}^\p}{\psi_T}}{\mel{\psi_T}{\mathcal{G}^\p \mathcal{G}^\p}{\psi_T}} .\label{eq:obs_general}
\end{align}
This expectation value can be stochastically sampled: For observables that are diagonal in the computational basis $\mathcal{O}_{xx'} \equiv \mathcal{O}(x) \delta_{xx'}$ the procedure, called \emph{forward-walking}, is particularly simple and outlined in more detail in Ref.~\cite{Buonaura1998}.
Each step, we keep track of the weights $w_{x_n}$, defined via \cref{eq:prob_weights}. From these, we compute the \emph{accumulated weight}
\begin{equation}
    W^{\p}_n = \prod_{i = n-\p}^{n+\p} w_{x_i}\label{eq:accum_weight},
\end{equation}
which gives a stochastic estimator for a projection for $\p$ steps each before and after step $n$.
Noting that applying $\mathcal{O}$ at step $n$ does not alter the configuration but only multiplies it by a number $\mathcal{O}(x_n)$, we obtain from \cref{eq:obs_general}
\begin{align}
   \langle \mathcal{O} \rangle =\lim_{\p \rightarrow \infty} \frac{\sum^{N-\p}_{n=\p} W^{\p}_{n} \mathcal{O}(x_{n})}{\sum^{N-\p}_{n=\p} W^{\p}_n}.\label{eq:GFMCobservable}
\end{align}
To compute the expectation value of the energy, we can use that $\mathcal{H}$ commutes with $\mathcal{G}$ so that we only need to replace $\mathcal{O}(x_n)$ by the \emph{local energy} $e_L(x_n) = \sum_{x'}\mathcal{H}_{x_nx'}$ in \cref{eq:GFMCobservable}.
Note that this is equivalent to the variational energy of $\mathcal{G}^{\p}\ket{\psi_T}$, whose expectation value must be strictly greater or equal to the ground state energy (up to statistical errors).

For general offdiagonal observables, applying $\mathcal{O}$ changes the configuration $\ket{x_n}$, so that a new simulation of length $\p$ needs to be run starting from one of the possible configurations contained in $\mathcal{O} \ket{x_n}$ (chosen with weights given by $\mathcal{O}_{x'x_n}$ \footnote{Or, in the case of importance sampling introduced later, $\mathcal{O}_{x'x_n} \psi_T(x')/\psi_T(x_n)$}).

The above procedure converges to the correct mean, however, the product in \cref{eq:accum_weight} leads to strong fluctuations which diverge exponentially in $\p$ rendering this basic approach extremely ineffective. In order to reduce the statistical fluctuations, two important additions to the formalism are necessary. 

\emph{Importance sampling --}
Clearly, the projection procedure would be improved by using a trial wavefunction $\ket{\psi_T}$, which is closer to the ground state.
Likewise, the sampling via the equilibrium condition above is a rather inefficient way to explore the Hilbert space. Better convergence can be achieved by favoring configurations that are expected to have a large overlap with the ground state wavefunction. 

Clearly, by modifying the projector as $\mathcal{G}_{xx'} \rightarrow \tilde{\mathcal{G}}_{xx'} = \mathcal{G}_{xx'} \times \psi_G(x')/\psi_G(x)$, the equilibrium distribution will be changed. Choosing $\psi_G(x) > 0$, we will thus sample more configurations that have a large overlap with $\ket{\psi_G}$. Due to this property, this function is known as a \emph{guiding wavefunction}. 

For $\psi_T(x) \neq 1$, the expectation values of diagonal operators only change implicitly via the modified weights $\tilde{w}_x =\sum_{x'} \tilde{\mathcal{G}}_{xx'}$, while the local energy must be modified as $e_L(x_n) =  \sum_{x'}\mathcal{H}_{x_nx'} \psi_G(x')/\psi_G(x_n)$, as the offdiagonal elements of the projector $\tilde{\mathcal{G}}$ are no longer proportional to $\mathcal{H}$.

It is found that if $\psi_G$ is chosen to be the exact ground state wavefunction, the variance of the energy is exactly zero and the projection converges at $\p = 0$. In the previous example, this was done by implicitly selecting $\psi_G(x) =1$. In the generic case, the exact solution is hard to find (otherwise there would be no reason to use GFMC after all). However, the required number of projection steps and statistical fluctuations can still be greatly reduced by choosing a good variational function.

While this function can be arbitrary in principle, practicality requires that $\braket{x}{\psi_G}$ must be efficient enough to evaluate for any given spin configuration.
Here, we use a simple Jastrow ansatz which is efficient to evaluate and captures two-body correlations to arbitrary length
\begin{equation}
    \log \psi_G(x) = \sum_{i=1}^{N_\textrm{sites}}{m_i x_i} + \sum_{i<j}{V_{ij} x_ix_j}.
\end{equation}
Here, we defined $x_i \equiv S_i^z$ and $m_i$ and $V_{ij} = V_{ji}$ are variational parameters that are optimized using the stochastic reconfiguration method~\cite{Sorella1998_recon} for each $\mu$ before the GFMC simulation. It is also possible to choose even more sophisticated (but costly) variational approaches, such as neural network states or even results from tensor network methods \cite{Carleo2017,He-Yu2024}. 
Finally, we stress that the guiding wave function itself is variational in nature and thus may or may not accurately represent the ground state properties. However, any bias introduced in this choice is eliminated by the projection procedure provided that the guiding wavefunction is nonsingular, i.e. $0 
\leq \psi_G(x) \neq \infty$ for any $x$. 

\emph{Many walker formalism--}
If the guiding wavefunction is not exact, the accumulation of weights still leads to exponentially increasing fluctuations at large $\p$. To alleviate this problem, one generally relies on an alternative formalism with several walkers. Here, we use the formalism following Ref. \cite{Buonaura1998}, which is an improvement of earlier approaches \cite{Runge1992,Hetherington1984}. The many-walker approach is used as follows:
First, $N_w$ walkers $\alpha = 1,\dots N_w$ are initialized and evolve fully independently for $n_\textrm{branch}$ steps, accumulating their weights $w_{\alpha, n}$ as $w_\alpha = \prod^{n_\textrm{branch}}_{n=1} w_{\alpha, n}$. Afterwards, we perform the reconfiguration in which each walker is given a new configuration from all walkers, sampled with probability $w_\alpha/\sum^{N_w}_{\alpha=1} w_\alpha$. This process is interpreted as a single Markov step of a new kind of Markov chain for an \emph{ensemble} of configurations with a single weight $w =\frac{1}{N_w}\sum^{N_w}_{\alpha=1} w_\alpha $. 
This reconfiguration ensures that walkers with negligible weight are eliminated while those with large weight multiply which reduces the population in irrelevant regions of the Hilbert space.
We note that to measure observables with the forward walking method, the procedure needs to be adapted to utilize the exact reconfiguration process, as some of the walkers are eliminated during the reconfiguration steps following the application of $\mathcal{O}$. See Refs.~\cite{Buonaura1998, becca2017quantum} for more details.

\emph{Continuous time limit --} For finite values of $\mu$, the value of $\Lambda$ needs to become very large to guarantee a non-negative projector. For most configurations, this results in a large value of $\mathcal{G}_{xx}$, implying a large probability to remain in the same state. For efficiency, we thus directly employ the continuous time limit~\cite{Ralko2005,ralkoDynamicsQuantumDimer2006}, which performs the limit $\Lambda\rightarrow\infty$, in which case the projector becomes an imaginary time evolution operator $\exp(-\mathcal{H}\Delta\tau)$. Here, increasing the imaginary time step $\Delta\tau$ corresponds to propagating the walkers for more steps between reconfigurations, analogous to increasing $n_\mathrm{branch}$. Note that the choice of $\Delta \tau$ is not affected by trotterization errors as the imaginary time projector is implemented exactly. However, an excessively large time step generally results in a small survival rate of the walkers which decreases sampling efficiency. For further details, we refer to Ref.~\cite{becca2017quantum}.

\emph{Ergodicity --}
In some cases, the Markov chain underlying the  GFMC procedure can be affected by poor ergodicity for example if the graph of the Hamiltonian features several clusters of states with relatively large weights $w_x$ which are separated by many states with much smaller weights \cite{Ralko2005,Sikora2011}. In this case, it may be statistically rare to transition from one such cluster to another. We diagnose this problem by initializing the walkers with a random configuration in the same subsector, by performing $\sim 100{,}000$ fully random cluster flips for each walker at initialization of the simulation. Statistical errors are then estimated by repeating this procedure for several independent runs. If ergodicity is broken, this typically results in large standard deviations of the mean energy resulting from walker ensembles becoming stuck in different clusters. In the present work, the system evolves ergodically in any considered sector for the entire range $0 \leq \mu \leq J'$.
\begin{figure}
    \centering
    \includegraphics[width=\linewidth]{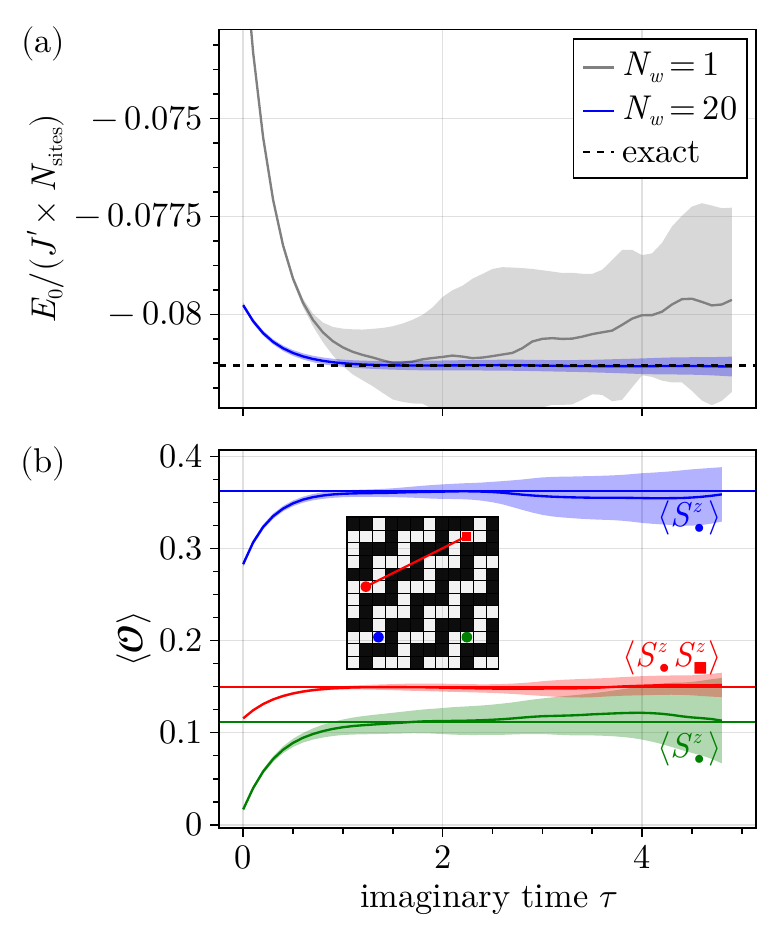}
    \caption{Comparison of GFMC in the spin-$1/2$ staircase sector with ED for $L=12$ with open boundary conditions and $\mu=0$. (a) Ground state energy as a function of the projection time for a single walker (grey) compared to an ensemble of 20 walkers (blue) and the exact result (black dashed line). (b) shows different expectation values of different one- and two-point spin correlators shown in the inset obtained from 20 walkers, each compared to the exact result by a solid line in the same color.}
    \label{fig:GFMCdemo}
\end{figure}

A demonstration of the application of GFMC is given in \cref{fig:GFMCdemo} for the staircase sector at $L=12$ for open boundaries and $\mu=0$. For this size, we may compare our results to exact diagonalization (ED) within this subsector of the Hilbert space, spanning 12522 configurations. A good indicator for the convergence of the projection is the ground state energy shown in \cref{fig:GFMCdemo}(a). The data point at $\tau = 0$ for the single walker corresponds to a variational Monte Carlo (VMC) estimate using the trial wavefunction (here, a Rokhsar-Kivelson function $\psi_G(x)=1$, which is far from the ground state at $\mu=0$). The many-walker formalism can be seen to improve statistical errors, particularly at higher projection orders. In \cref{fig:GFMCdemo}(b) we show how arbitrary observables in $S^z$ basis can be measured using the forward walking formalism. All observables can be seen to be converged to the exact result after $\tau \approx 1.75$ for $N_w=20$ walkers. For the projection, we took $\Delta \tau =0.1$, and used $50{,}000$ Monte Carlo steps for the $N_w=20$ result, while for $N_w = 1$, this number was increased by a factor of 20 for a more direct comparison.

Generally, the required projection usually depends on the system size $L$ and the number of walkers. In the present case we find that the relatively small size of individual Hilbert space sectors significantly reduces numerical overhead.
For example, for \cref{fig:SpinHalfOverview}(e), we found excellent convergence of the projection for an ensemble of $300$ walkers evolving under the continuous time formalism. At the RK point, the guiding wavefunction is exact, which further reduces the effort, as no projection is necessary and a single walker is sufficient (though larger numbers were used here in order to parallelize the sampling). 
The mean and standard deviations (error bars) were estimated by performing $12$ independent runs. 

The code used here was implemented in the Julia language and can be found at \url{https://github.com/NilsNiggemann/SpiderWebModel.jl}.

\section{Exhaustive search of sectors}\label{app:all_sectors}
\Cref{fig:SpinHalfOverview}(h) displays the energy of several sectors with the highest density of flippable clusters. While the density of flippable clusters plays an important role in a sector's energy, it is in principle possible that sectors exist which feature many configurations where flipping a cluster does not result in a reduction of the number of flippable clusters. If they exist, such sectors could possibly have energies that are competitive with the energy of the staircase sector, especially for large $\mu \lesssim 1$.
In order to systematically rule out the presence of such sectors, we performed an exhaustive search of periodic fracton-free configurations with varying unit cells $L_x \times L_y$, where $L_x\leq 6$ and $L_y \leq 6$. We allowed for all periodic tilings of these unit cells, by defining the tiling's lattice vectors as $(L_x,o)$ and $(0,L_y)$, where $o \in \mathds{Z}, 0 \leq o < L_y$ is an offset.
Excluding sectors with no flippable clusters, we find no more than $105$ inequivalent Hilbert space sectors that are distinct under time-reversal, translation, rotation, and mirror symmetries~\footnote{Note that this approach allows for inequivalent sectors that can still have equivalent quantum dynamics if the arrangement of flippable clusters is equivalent.}. Using GFMC, we then compute the ground state energy in each of these sectors to identify the lowest-lying ones. To minimize the impact of finite size effects, we consider systems of size $18\times18$ and larger, ensuring that each system size is a multiple of the \emph{square} unit cell size (which may exceed $6\times 6$ for non-square tilings).

\begin{figure}
    \centering
    \includegraphics[width=\linewidth]{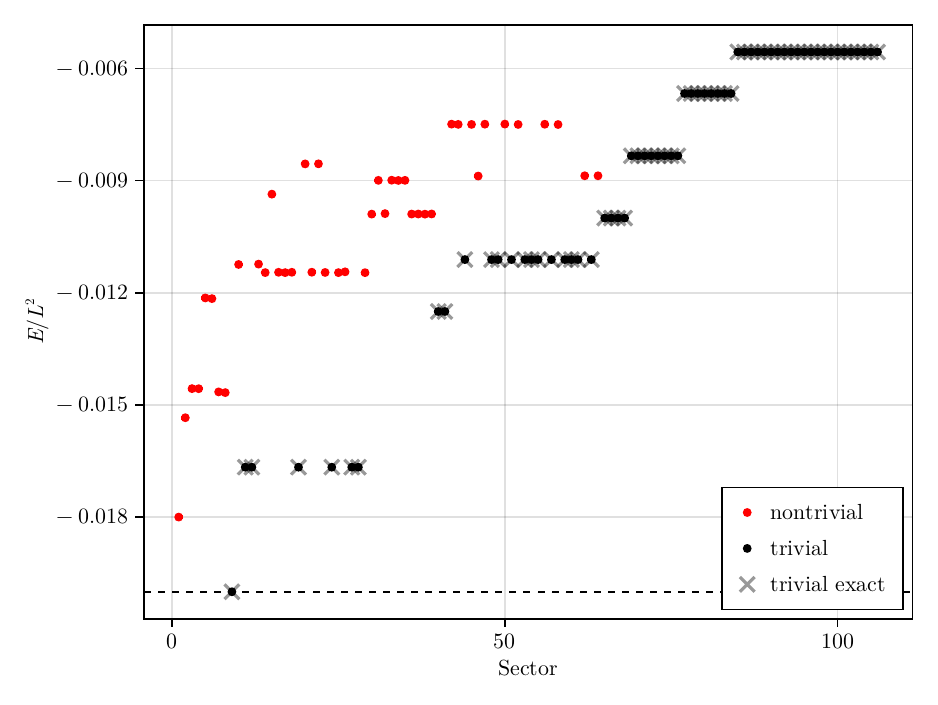}
    \caption{Energies of sectors defined via periodic configurations with unit cells up to $6\times6$ for $\mu = 0.8J'$. Sectors are sorted according to their maximal number of flippable clusters. Red markers indicate sectors with non-trivial quantum dynamics where flippable clusters overlap. Exact energies for trivial sectors are computed via \cref{eq:exact_trivial}.}
    \label{fig:spin_half_sectors_ens_mu08}
\end{figure}

\Cref{fig:spin_half_sectors_ens_mu08} shows the energies obtained this way for a value of $\mu=0.8J'$. The sectors are sorted according to their maximum degree, i.e., the maximum number of flippable clusters for the maximally flippable configuration, such that sector number one corresponds to the staircase sector. We further indicate trivial sectors, where any cluster flip neither adds nor disables other flippable clusters, via black markers. The energy of these sectors can be computed exactly as 
\begin{equation}
    E = -(J'-\mu)N_{\sublO} \label{eq:exact_trivial},
\end{equation}
under the assumption that combinations of the $N_{\sublO}$ possible cluster flips likewise do not affect any other clusters. The perfect agreement between numerical and exact results confirms that this assumption is never violated.

Despite the large RK-potential, the lowest energies are given by trivial sectors, with the only exception being the staircase sector. This finding supports the conclusion that no other sector is capable of supporting a true spin liquid ground state.

\end{appendices}
\end{document}